\documentclass[longauth]{aa}  

\usepackage{graphicx}
\usepackage{txfonts}
\usepackage{lipsum}
\usepackage{subcaption}         % necessary for continued figures, example in section 3
\usepackage{lscape}             % to rotate a single page table, example in appendix.
\usepackage{placeins}           % useful with \FloatBarrier, to keep 
\usepackage{wrapfig}
\usepackage{floatrow}

\usepackage[colorlinks=true,linkcolor=blue,citecolor=blue,urlcolor=blue]{hyperref}%
\usepackage[dvipsnames]{xcolor}
\usepackage{upgreek}

\newcommand{\ciii}{\ion{C}{iii}}
\newcommand{\hi}{\ion{H}{i}}

\newcommand{\lya}{Ly$\alpha$\xspace}
\newcommand{\oiii}{O\,{\sc iii}}

\newcommand{\jwst}{\emph{JWST}\xspace}

\begin{document}

   %\title{Damped Ly$\alpha$ absorbers reveal efficient star formation and hidden redshift biases in the first galaxies at $z>10$} 
   \title{JWST spectroscopy of galaxies at $z>10$: Damped Ly$\alpha$ absorbers reveal efficient star formation and hidden redshift biases}
   % JWST spectroscopy of galaxies at z>10: damped Lyα absorbers reveal dense gas reservoirs, efficient star formation, and redshift biases
   % Damped Lyα absorption in JWST spectra of z>10 galaxies: probing gas surface densities, star formation efficiencies, and redshift biases
   % Tracing the first galaxies with JWST spectroscopy: damped Lyα absorbers as probes of gas surface density and redshift systematics at z>10
   %\title{Evidence for efficient star formation occurring in dense, neutral gas reservoirs of nascent UV-bright galaxies at $z>10$} % In the first 500 Myr / at cosmic dawn

  %  \subtitle{Subtitle}

%%%%%%%%%%%%%%%%%%%%%%%%%%%%%%%%%%%%%%%%
% Please do not include ORCIDs next to author names.
% Only ORCIDs authenticated by individual authors in EDP Sciences editorial system will be taken into account.
% ORCIDs included here will be removed.
%%%%%%%%%%%%%%%%%%%%%%%%%%%%%%%%%%%%%%%%

   \author{Kasper~E.~Heintz\inst{1,2,3}
   \and Clara~L.~Pollock\inst{2,3}
   \and Joris~Witstok\inst{2,3}
   \and Rychard~J.~Bouwens\inst{4}
   \and Sandro~Tacchella\inst{5,6}
   \and Pascal~A.~Oesch\inst{7,2,3}
   \and Pratika~Dayal\inst{8,9,10}
   \and Sownak~Bose\inst{11}
   \and Gabriel~B.~Brammer\inst{2,3}
   \and Johan.~P.~U.~Fynbo\inst{2,3}
   \and Rashmi~Gottumukkala\inst{2,3}
   \and Matthew~J.~Hayes\inst{12}
   \and Akio~K.~Inoue\inst{13,14}
   \and Benjamin~Johnson\inst{15}
   \and Harley~Katz\inst{16,17}
   \and Peter~Laursen\inst{2,3}
   \and Rohan~P.~Naidu\inst{18,19} 
   \and Desika~Narayanan\inst{20,2,3}
   \and Lucie~E.~Rowland\inst{4}
   \and Nial~R.~Tanvir\inst{21}
   \and Chamilla~Terp\inst{1,2,3}
   \and Sune~Toft\inst{2,3}
   \and Francesco~Valentino\inst{1,2}
   \and Fabian~Walter\inst{22,23}
   \and John~R.~Weaver\inst{24}
   \and \\ Arjen~van~der~Wel\inst{25}
   \and Mengyuan~Xiao\inst{7}
   }

   \institute{DTU Space, Technical University of Denmark, Elektrovej 327, DK2800 Kgs. Lyngby, Denmark; \\
    \email{kelhe@space.dtu.dk}
    \and Cosmic Dawn Center (DAWN), Denmark
    \and Niels Bohr Institute, University of Copenhagen, Jagtvej 128, 2200 Copenhagen N, Denmark
    \and Leiden Observatory, Leiden University, NL-2300 RA Leiden, Netherlands
    \and The Kavli Institute for Cosmology (KICC), University of Cambridge, Madingley Road, Cambridge, CB3 0HA, UK
    \and Cavendish Laboratory, University of Cambridge, 19 JJ Thomson Avenue, Cambridge, CB3 0HE, UK
    \and Department of Astronomy, University of Geneva, Chemin Pegasi 51, CH1290 Versoix, Switzerland
    \and Canadian Institute for Theoretical Astrophysics, 60 St George St, University of Toronto, Toronto, ON M5S 3H8, Canada
    \and David A. Dunlap Department of Astronomy and Astrophysics, University of Toronto, 50 St George St, Toronto ON M5S 3H4, Canada
    \and Department of Physics, 60 St George St, University of Toronto, Toronto, ON M5S 3H8, Canada
    \and Institute for Computational Cosmology, Department of Physics, Durham University, South Road, Durham DH1 3LE, UK
    \and Stockholm University, Department of Astronomy and Oskar Klein Centre for Cosmoparticle Physics, AlbaNova University Centre, SE-10691, Stockholm, Sweden
    \and Waseda Research Institute for Science and Engineering, Faculty of Science and Engineering, Waseda University, 3-4-1 Okubo, Shinjuku, 169-8555 Tokyo, Japan
    \and Department of Physics, School of Advanced Science and Engineering, Faculty of Science and Engineering, Waseda University, 3-4-1 Okubo, Shinjuku, 169-8555 Tokyo, Japan
    \and Center for Astrophysics | Harvard \& Smithsonian, 60 Garden St., Cambridge, MA 02138, USA
    \and Department of Astronomy \& Astrophysics, University of Chicago, 5640 S Ellis Avenue, Chicago, IL 60637, USA
    \and Kavli Institute for Cosmological Physics, University of Chicago, Chicago IL 60637, USA
    \and MIT Kavli Institute for Astrophysics and Space Research, 70 Vassar Street, Cambridge, MA 02139, USA
    \and Institute for Astronomy, University of Hawai’i, 2680 Woodlawn Drive, Honolulu, HI 96822, USA
    \and Department of Astronomy, University of Florida, 211 Bryant Space Sciences Center, Gainesville, FL 32611 USA
    \and School of Physics and Astronomy, University of Leicester, University Road, Leicester, LE1 7RH, United Kingdom
    \and Max Planck Institut f\"ur Astronomie, K\"onigstuhl 17, D-69117 Heidelberg, Germany
    \and California Institute of Technology, Pasadena, CA 91125, USA
    \and MIT Kavli Institute for Astrophysics and Space Research, 70 Vassar Street, Cambridge, MA 02139, USA
    \and Department of Physics and Astronomy, Universiteit Gent, Proeftuinstraat 86 N3, B-9000 Ghent, Belgium
    }

   \date{Received \today}

% \abstract{}{}{}{}{}
% 5 {} token are mandatory
 
    \abstract{
    Recent observations with \jwst\ have revealed a remarkable population of surprisingly luminous galaxies within the first 500\,Myr of cosmic time at redshifts $z>10$.
    %, both in terms of their abundance and brightness at surprisingly early epochs. 
    Their abundance exceed predictions from simulations and empirical extrapolations from lower redshifts, suggesting a transition in the physical conditions under which the first stars formed. Here we investigate the physical conditions of a select sample of 25 galaxies with robust redshift measurements at $z_{\rm spec}\geq 10$ observed with \jwst/NIRSpec Prism. We characterize their star-formation efficiency, `burstiness', and presence of strong rest-frame UV nebular lines in relation to the density of the local neutral atomic hydrogen (\hi) gas reservoirs they are embedded in. We find that the prominence of strong rest-UV lines are correlated with the burstiness of the galaxies, defined as ${\rm SFR_{10\,Myr} / SFR_{100\,Myr}}$. In contrast, there are no strong connections between the \hi\ gas column density derived from the damped \lya\ absorption (DLA) and the $M_{\rm UV}$ brightness, ${\rm SFR_{10\,Myr} / SFR_{100\,Myr}}$, and prominence of rest-UV lines. We find evidence for the most bursty galaxies showing a large variation in star-formation efficiencies and \hi\ gas surface densities, though typically with very short depletion timescales, $t_{\rm dep} \lesssim 20$\,Myr. This necessites rapid gas depletion times and external replenishment from infalling, pristine gas, powering starburst episodes on equally short timescales. We further quantify the impact of strong DLAs in galaxy spectra on photometric and \lya-break redshift-inferences, finding average redshift biases of $\langle z \rangle =0.39$ and $0.14$, respectively, when not incorporating DLAs on the emergent spectra. We show the effect of this bias on new measurements of the cosmic UV luminosity density, $\rho_{\rm UV}$, derived here at $z>10$, finding that this has a marginal impact on the UV luminosity function. 
    In conclusion, we provide observational evidence for highly efficient star formation occurring in dense gas with rapid depletion timescales for galaxies at $z>10$, and highlight the perils of not accounting for strong Ly$\alpha$ absorption in photometric or Lyman-break redshift determinations. 
    }


    %\titlerunning{Dense gas reservoirs power UV-bright galaxies at $z>10$}
    \titlerunning{Efficient star formation powered by dense, neutral gas reservoirs in UV-bright galaxies at $z>10$}
    \authorrunning{Heintz et al.}

   \maketitle

\section{Introduction}
\label{sec:intro}

The first stars and galaxies are believed to emerge from the `dark ages' of the Universe at redshifts $z\approx 20-30$ \citep{Dayal18,Robertson22,Stark25,Bouwens2026}, heralding cosmic dawn. This process is initiated by the gravitational infall and collapse of primordial neutral atomic hydrogen (\hi) onto dark matter halos, which serve as the primary fuel for the formation of the first stellar sources. With the recent advent of powerful space-based observatories, in particular the {\em James Webb Space Telescope} \citep[\jwst,][]{Gardner06}, we are now starting to approach this redshift frontier, with robust spectroscopic measurements of luminous galaxies at $z\gtrsim 14$ \citep{Carniani24a,Naidu25,Wu25} and photometric measurements extending potentially out to $z\sim 15-25$ \citep[e.g.,][]{Robertson24,Whitler25,Weibel25,PerezGonzalez25,Castellano25a}.  

One of the perhaps most surprising discoveries in early \jwst observations was the apparent overabundance of UV-bright galaxies at $z\gtrsim 10$ \citep[e.g.][]{Castellano22, Naidu22, Atek23a, Bouwens23, Donnan23, Finkelstein23, Harikane23, Adams24, McLeod24, Robertson24, Carniani24a}, far greater in number and luminosity than predicted from (mostly pre-\jwst) simulations and extrapolations from the lower-redshift galaxy population. Several physical scenarios have been proposed to explain this, including extremely stochastic or `bursty' star-formation histories \citep{Sun23,Mason23,Shen23,Gelli24,Kravtsov24,Endsley25,Munoz26}, feedback-free starburst episodes \citep{Dekel23,Li24}, increased star-formation efficiencies \citep[][]{Somerville25,Mauerhofer25,Pollock26}, differences in the dust attenuation or the entire removal of dust from the galaxies \citep{Ferrara23,Mckinney25}, higher UV-light to mass ratios driven by a more pronounced `top-heavy' stellar initial mass function \citep[IMF;][]{Cowley18,Inayoshi22,Yung24,Hutter25,Mauerhofer25,Lu25}, strong nebular emission continuum \citep{Cameron24,Katz25,Reumert26}, or substantial contributions from active galactic nuclei \citep[AGN, e.g.,][]{Hedge24}. Fundamentally, the majority of these scenarios imply a transition in the primary mode of star formation in galaxies within the first 500\,Myr of cosmic time.  

The spectroscopic \jwst\ legacy samples are now starting to become statistically significant at $z>10$, enabling population analyses of the physical properties of galaxies at this early epoch. The first detailed spectroscopic characterization of galaxies at $z\gtrsim 10$ has indeed revealed a remarkable large diversity in physical properties, with some galaxies showing strong rest-frame UV emission lines \citep{Bunker23_gnz11,CurtisLake23,Calabro24b,Castellano25b,Naidu25}, and others with a near opposite dearth of lines \citep{ArrabalHaro23,Carniani24a,Witstok25,Wu25,Donnan26}. This potentially hints at a dichotomy in ionization potential and chemical abundance patterns \citep{Harikane24,Cameron23b} or large variations in the ionizing photon escape fractions \citep{MarquesChaves26}. Generally, galaxies at this epoch are observed to be relatively metal-poor with oxygen abundances $2-3\times$ lower than expected from their star-formation rates (SFRs) and stellar masses \citep{Pollock25}. Further, they show more prominent and prevalent nebular emission lines \citep{Tang25}, hinting at more rapid star-formation timescales or a large diversity in the occurrence between bursts and lulls of star formation \citep{RobertsBorsani25}. Finally, the majority of galaxies at $z>10$ show broad damped \lya\ absorption \citep[][]{Heintz24_DLA,Heintz25_z14,Umeda23,Carniani24a,Pollock26}, indicating the presence of extremely dense \hi\ gas in the local surroundings of these galaxies \citep{Rowland25,Gelli25}. 

We here seek to resolve the underlying physical mechanism that drives the remarkable UV luminosity of galaxies at $z>10$ using a similar statistical sample of 22 galaxies, all spectroscopically confirmed to be at $z>10$, and with high-quality spectroscopic and photometric data. We investigate two potential scenarios that could either drive or mimic extreme UV brightness in this early galaxy population: {\it i)} recent elevated star-formation efficiencies and {\it ii)} overestimated distances and intrinsic luminosities. Specifically, we aim to understand the impact of the dense, local \hi\ gas on driving the observed star-formation properties and the induced bias from strong DLAs on the derived redshifts when only information from photometry or the \lya\ break in spectroscopy is available. The latter point is particularly important at $z>10$ where most of the prominent rest-frame optical nebular lines are redshifted out of the \jwst/NIRSpec wavelength coverage. In the companion paper by \citet{Pollock26}, we have already shown that galaxies at these redshifts generally have high star-formation efficiencies (SFEs) or short gas depletion timescales based on their offset from the local `Kennicutt-Schmidt' relation \citep{Kennicutt12}. Here, we focus on deriving the impact of the gas density itself on the burstiness and SFE of this early galaxy population.  

We have structured the paper as follows. In Sect.~\ref{sec:data}, we present the compilation of \jwst\ data used in this work and detail the spectro-photometric analysis of each source, including a careful evaluation of the derived redshifts inferred from photometry, the Lyman-$\alpha$ break, and strong nebular emission lines. Sect.~\ref{sec:res} presents the results on the inferred star-formation efficiency and `burstiness' of the target galaxies. Finally, in Sect.~\ref{sec:uvlf} we quantify the redshift bias when considering only the photometric or \lya-break inferences and show the implications of this bias on a new constructed UV luminosity function curve here and compared to the literature. In Sect.~\ref{sec:conc} we summarize and conclude on our work. 

Throughout the paper, we adopt a concordance flat $\Lambda$CDM cosmological model, assuming a Hubble constant $H_0 = 67.4$\,km\,s$^{-1}$\,Mpc$^{-1}$ and matter and vacuum energy density parameters $\Omega_{\rm m} = 0.315$ and $\Omega_{\rm \Lambda} = 0.685$ \citep{Planck18}.

\section{Observations and analysis}
\label{sec:data}

\subsection{Observations and $z>10$ sample}

The spectroscopic data entering the main sample analyzed in this work is extracted from the DAWN \jwst\ archive (DJA)\footnote{\url{https://dawn-cph.github.io/dja/}}, including all publicly available spectra taken with \jwst\ to date. We only consider the spectra taken with the \jwst/NIRSpec Prism configuration \citep[][]{Jakobsen22}, using the data products from the fourth DJA version \citep[v4, see][]{Valentino25,Pollock25} building on previous releases \citep{Brammer_msaexp,Heintz25,DeGraaff25}. Notably, this new version includes a bar shadow correction, greatly improving the absolute and color-dependent flux calibration, and extends the nominal wavelength range up to $0.6-5.5\mu$m at a spectral resolving power of $\mathcal{R}=30-300$. The delivered spectral resolutions are typically higher by $30-50\%$ \citep{DeGraaff24,Pollock25}, depending on the size of the source and its location within the shutter. We compile all spectra from the DJA with secure spectroscopic redshifts $z\ge 10$ and with ${\rm S/N}>3$ per wavelength bin in the rest-frame UV. We include new data for two galaxies at $z>10$ from the {\em `Mirage or Miracle'} (MoM) survey (Prog. ID = 5224, PIs: Oesch \& Naidu), two of them already presented as MoM-$z14$ and MoM-$z11$ in \cite{Naidu25} and \cite{RobertsBorsani25}. 

\begin{figure*}[!h]
    \centering
    \includegraphics[width=\linewidth,height=21cm]{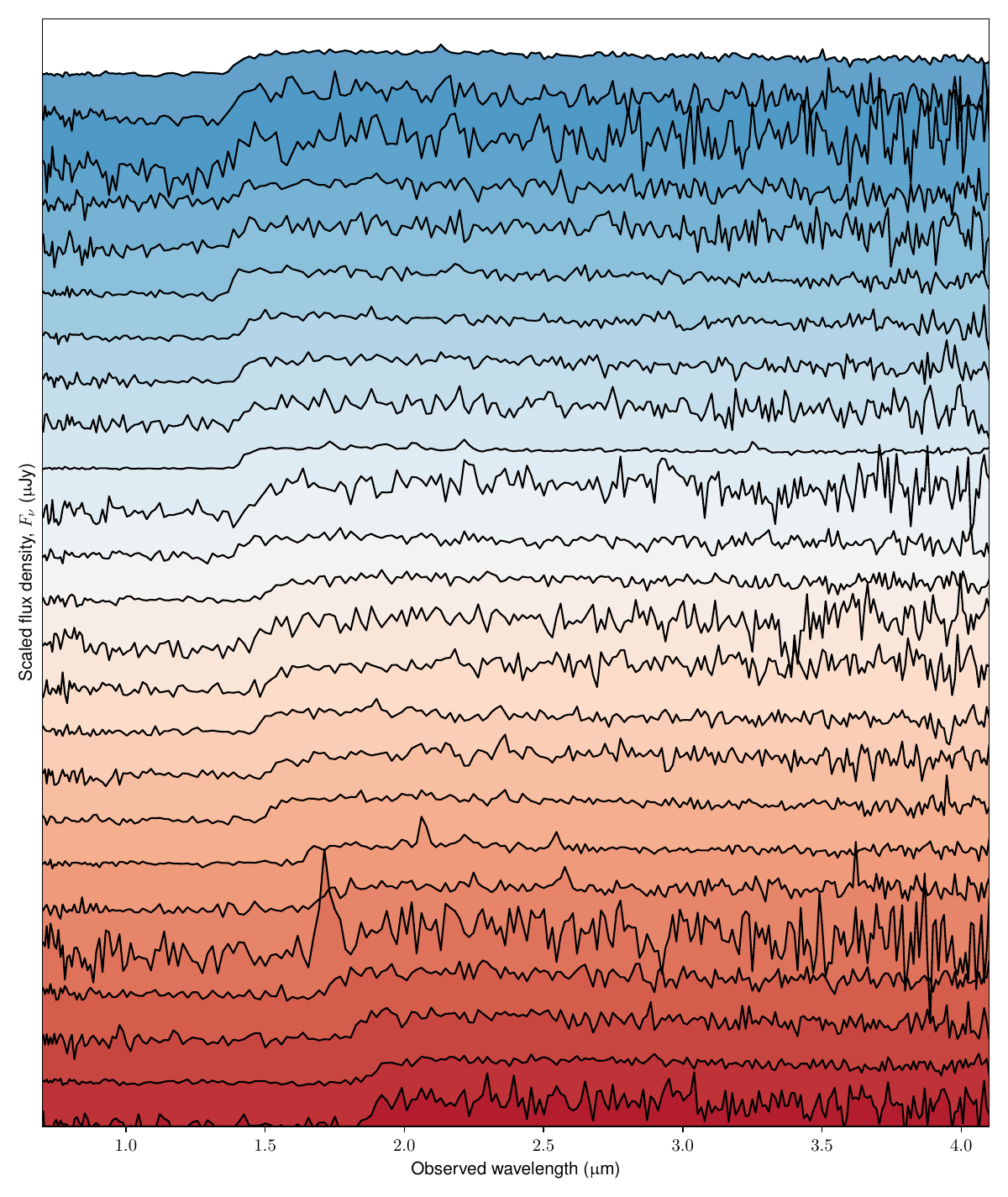}
    \caption{Compilation of galaxy spectra at $z>10$. All spectra are observed with \jwst/NIRSpec in the Prism configuration ($\mathcal{R}\sim 100$, $\lambda = 0.6-5.5\upmu$m). The sources are shown from lowest (top) to highest (bottom) redshifts and are all selected to have ${\rm S/N}>3$ per wavelength bin in the rest-frame UV (at $\approx 1500\,\AA$), enabling detailed modeling of the \lya\ breaks redshifted out to $\sim 1.3-1.9\upmu$m at $z=10.0-14.4$. The exact redshifts and source IDs are summarized in Table~\ref{tab:tab1}.}
    \label{fig:spectra}
\end{figure*}

The full set of \jwst/NIRSpec Prism spectra for the compiled galaxies at $z\ge 10$ considered in this work are shown in Figure~\ref{fig:spectra}, visualizing their similarities and diversities, and their source properties are summarized in Table~\ref{tab:tab1}, including references to the original survey papers or previous works on the same sources in the literature. 
% and described in detail for each source in the Appendix. -- TBD

%\clearpage

%%% TABLE %%%%
\begin{table*}
\begin{center}    % used for centering table
\caption{Target IDs, coordinates, redshifts, and absolute UV brightness for the compiled galaxies at $z>10$.} % title of Table
\label{tab:tab1}      % is used to refer this table in the text
\setlength\tabcolsep{0.2cm}
\renewcommand{\arraystretch}{1.2}
\begin{tabular}{c c c c c c c c}
\hline\vspace{0.1cm}%\hline                        % inserts double horizontal lines
Source ID & R.A. & Decl. &  $z_{\rm spec}$ & $z_{\rm break}$ & $z_{\rm phot}$ & $M_{\rm UV}$ & Ref. \\ % table heading
& (deg) & (deg) & & & & (mag)  & \\
% (deg) & (deg) & & & & \\
 (1) &  (2) &  (3) &  (4) & (5) & (6) & (7) & (8)  \\
\hline   % inserts single 
MoM-z14         & 150.09333 & 2.27316 & 14.4441 & $14.39^{+0.08}_{-0.11}$ & $14.86^{+0.47}_{-1.50}$ & $-19.86\pm 0.15$ & (1) \\
JADES-GS-z14-0  & 53.08294 & $-27.85563$ & 14.1832 & $14.44^{+0.04}_{-0.06}$ & $14.51^{+0.27}_{-0.28}$ & $-20.96\pm 0.09$ & (2,3) \\
JADES-GS-z14-1  & 53.07427 & $-27.88592$ & 14.0848 & $14.07^{+0.07}_{-0.08}$ & $14.36^{+0.82}_{-1.40}$ & $-19.32\pm 0.08$ & (2) \\
JADES-GS-z13-0  & 53.14988 & $-27.77650$ & 12.8479 & $13.21^{+0.07}_{-0.09}$ & $13.41^{+0.65}_{-0.49}$ & $-18.63\pm 0.04$ & (4,5) \\
JADES-GS-z13-1-LA & 53.06475 & $-27.89024$ & 12.7959 & $--$                  & $12.93^{+0.08}_{-0.16}$ & $-19.00\pm 0.40$ & (6) \\
JADES-GS-z12-0  & 53.16635 & $-27.82156$ & 12.4978 & $12.89^{+0.08}_{-0.09}$ & $13.0^{+0.5}_{-0.5}$    & $-18.98\pm 0.04$ & (4,5,7) \\
GHZ2/GLASS-z12  & 3.49898 & $-30.32475$ & 12.3420 & $12.49^{+0.04}_{-0.05}$ & $12.22^{+0.13}_{-0.17}$ & $-21.22\pm 0.04$ & (8) \\
CAPERS-88619    & 34.31913 & $-5.23014$ & 11.3637 & $11.45^{+0.03}_{-0.02}$ & $--$                    & $-20.76\pm 0.04$ & \\
CEERS-16943     & 214.94314 & 52.94244 & 11.3408 & $11.56^{+0.08}_{-0.08}$ & $11.6^{+0.4}_{-0.5}$    & $-20.13\pm 0.06$ & (9) \\
JADES-GS-z11-1  & 53.11763 & $-27.88818$ & 11.2499 & $11.24^{+0.05}_{-0.05}$ & $11.2^{+1.0}_{-1.0}$    & $-19.09\pm 0.08$ & (10) \\
CEERS-10        & 214.90663 & 52.94550 & 11.1962 & $11.50^{+0.10}_{-0.09}$ & $10.95^{+0.35}_{-0.35}$ & $-20.00\pm 0.11$ & (11) \\
CAPERS-126973   & 34.26444 & $-5.09623$ & 11.1736 & $11.18^{+0.13}_{-0.15}$ & $12.5^{+0.9}_{-0.9}$    & $-20.21\pm 0.16$ &  \\
JADES-GS-z11-0  & 53.16477 & $-27.77463$ & 11.0979 & $11.43^{+0.07}_{-0.05}$ & $11.29^{+0.23}_{-0.18}$ & $-19.40\pm 0.04$ & (12) \\
CAPERS-22637    & 214.93207 & 52.84187 & 10.7457 & $10.76^{+0.04}_{-0.05}$ & $10.64^{+0.20}_{-0.30}$ & $-20.36\pm 0.09$ &  \\
MoM-z11         & 34.39431 & $-5.12917$ & 10.7123 & $11.09^{+0.10}_{-0.10}$ & $13.5^{+1.0}_{-1.3}$    & $-19.38\pm 0.17$ & (13) \\
GNz11           & 189.10605 & 62.24205 & 10.6296 & $10.62^{+0.03}_{-0.03}$ & $11.1^{+0.5}_{-0.5}$    & $-21.94\pm 0.01$ & (14) \\
GS-z10          & 53.14675 & $-27.79406$ & 10.5052 & $10.64^{+0.09}_{-0.10}$ & $--$                    & $-19.78\pm 0.11$ & (4) \\
CAPERS-136645   & 34.45602 & $-5.12195$ & 10.4980 & $10.61^{+0.05}_{-0.05}$ & $9.6^{+0.5}_{-0.5}$     & $-20.00\pm 0.05$ & (15) \\
JADES-20176151  & 53.07076 & $-27.86544$ & 10.4599 & $10.69^{+0.06}_{-0.08}$ & $10.7^{+0.3}_{-0.3}$    & $-19.37\pm 0.06$ &  \\
Abell-2744-22302 & 3.45137 & $-30.32073$ & 10.4597 & $10.44^{+0.03}_{-0.05}$& $10.67^{+0.19}_{-0.23}$ & $-20.14\pm 0.05$ & (16) \\
JADES-GS-z10-0  & 53.15884 & $-27.77349$ & 10.3989 & $10.32^{+0.08}_{-0.10}$ & $10.59^{+0.23}_{-0.24}$ & $-17.29\pm 0.29$ &  \\
Abell-2744-23984 & 3.45143 & $-30.32180$ & 10.3591 & $10.44^{+0.07}_{-0.07}$& $10.70^{+0.18}_{-0.2}$  & $-20.79\pm 0.09$ & (16) \\
CAPERS-82699    & 34.41356 & $-5.24425$ & 10.3960 & $10.43^{+0.18}_{-0.15}$ & $9.4^{+0.5}_{-0.5}$     & $-19.39\pm 0.15$ & \\
CAPERS-109917   & 150.14291 & 2.28802 & 10.2807 & $10.38^{+0.07}_{-0.08}$ & $10.5^{+0.5}_{-0.5}$    & $-19.41\pm 0.13$ & \\
MACS0647-3349   & 101.97133 & 70.23972 & 10.1716 & $10.47^{+0.03}_{-0.02}$ & $10.6^{+0.3}_{-0.3}$    & $-21.46\pm 0.05$ & (17) \\
\hline          %inserts single line
\end{tabular} 
\end{center}  
\textbf{Notes.} Column (1): Source ID / Name. Column (2): Right Ascension in degrees, J\,2000. Column (3): Declination in degrees, J\,2000. Column (4): Spectroscopic redshift from the identified emission-lines. The uncertainties are typically of the order $\Delta z = 0.02$. Column (5): \lya\ break redshift, assuming only an IGM-component. Column (6): Photometric redshift. Column (7): Absolute UV magnitude derived from the spectra, scaled to the match the photometry and corrected for lensing magnification. Column (8): Literature references for the individual galaxies. \\
{\bf References.} (1)~\citet{Naidu25}. (2)~\citet{Carniani24a}. (3)~\citet{Carniani25}. (4)~\citet{CurtisLake23}. (5)~\citet{Hainline24b}. (6)~\citet{Witstok25}. (7)~\citet{DEugenio24}. (8)~\citet{Castellano24}. (9)~\citet{Finkelstein23}. (10)~\citet{Scholtz26}. (11)~\citet{ArrabalHaro23}. (12)~\citet{Witstok25_O3}. (13)~\citet{RobertsBorsani25}. (14)~\citet{Bunker23}. (15)~\citet{Kokorev25}. (16)~\citet{Napolitano24}. (17)~\citet{Hsiao23}. 
%{\bf References.} ??
\end{table*}
%%%%%%%%%%%%%%

\subsection{Redshift estimates}

Since most of the prominent rest-frame optical nebular emission lines are redshifted out of the wavelengths covered by \jwst's NIRSpec's wavelength coverage at $z\gtrsim 10$, it is challenging to accurately constrain these high spectroscopic redshifts ($z_{\rm spec}$). For each galaxy, we thus carefully evaluate the rest-frame UV continuum and potential line emission, typically relying on the detection of \ciii]\,$\lambda 1909$ which is one of the strongest rest-frame UV lines detected at $z>10$ \citep[see][for more details]{RobertsBorsani25,Pollock26}. We use the redshifts inferred from photometry, $z_{\rm phot}$, compiled from the literature or DJA and the \lya\ break, $z_{\rm break}$, as priors on the upper bound of $z_{\rm spec}$ when searching for emission-line features, since these might overestimate the true systemic redshift due to prominent DLAs mimicking a higher redshift \lya\ break \citep[see e.g.,][]{ArrabalHaro23a,Fujimoto23_ceers,Finkelstein24,Heintz24_DLA,Heintz25,Hainline24b,Hainline26,Asada25,Witstok25_O3}. This will be discussed in more detail in Sect.~\ref{ssec:dla} below. 

Where available, we also include the systemic line redshift for the sample galaxies from the literature, when recovered, for instance, in targeted follow-up observations of the far-infrared (FIR) [\oiii]-$88\mu$m line emission with the Atacama Large Millimetre/submillimetre Array \citep[ALMA; see e.g.,][]{Witstok25_O3,Schouws24,Carniani25}. This accurately pin-points $z_{\rm spec}$ at FIR wavelengths (typically of order $\Delta z = 0.0001$) and enables a detailed characterization of the \lya\ damping wings, even when no prominent emission lines are detected in the rest-frame UV to optical \jwst\ spectra. The various redshift estimates from $z_{\rm spec}$, $z_{\rm break}$, and $z_{\rm phot}$, are summarized in Table~\ref{tab:tab1}. We will discuss and quantify the impact of strong DLAs on redshift estimates from $z_{\rm break}$ and $z_{\rm phot}$ in Sect.~\ref{sec:uvlf}, and the implications for global-derived galaxy properties such as the galaxy UV luminosity function when considering photometric measurements only.

\subsection{Damped Lyman-$\alpha$ absorption modeling} \label{ssec:dla}

For each of the galaxies at $z\geq 10$, we fit the observed \lya\ damping wings assuming two different physical components: \hi\ in the IGM along the line of sight and \hi\ in the local environment, following previous works \citep{Heintz24_DLA,Heintz25_z14,Pollock26}. Briefly, we describe the shape of the \lya\ damping wings as imprinted from the \hi\ in the IGM following the prescription by \citet{MiraldaEscude98}, though with the corrections by \citet{Totani06}. The DLAs representing the local dense \hi\ gas is modeled as Voigt profiles, following the approximation by \citet{TepperGarcia06}. We fit each of the compiled galaxy spectra using three different iterations:
\begin{itemize}
    \item {\bf IGM+ISM.} Our default model includes \lya\ damping wing absorption from a combination of a partly to fully neutral hydrogen fraction, $x_{\rm HI} \equiv n_{\rm HI}/n_{\rm H,tot}$, in the IGM along the line of sights and the column density of \hi\ gas, $N_{\rm HI}$, in the local environments of the galaxies (ISM or CGM). In this default model, we fix the DLA redshift and the upper bound on the IGM component to $z_{\rm spec}$. We assume this scenario to be the most common and physically motivated, since most of the galaxies observed at $z>10$ in the literature have implied \hi\ column densities greatly in excess of that imprinted from even a fully neutral IGM \citep[][though see \citealt{Mason26}]{Heintz24_DLA,Pollock26}. From this model, we extract the median and 16th to 84th percentiles from the posterior distributions of the main output parameters $x_{\rm HI}$ and $N_{\rm HI}$. This model thus provides the best constraints on the joint \hi\ absorption in the two physically distinct components. We caution, however, that $x_{\rm HI}$ will typically be degenerate with $N_{\rm HI}$ at low column densities ($N_{\rm HI}\lesssim 10^{21}$\,cm$^{-2}$), and is difficult to constrain at higher column densities to complete saturation from the DLA \citep[e.g.,][]{Huberty25}. 
    \item {\bf IGM only.} To gauge the bias on estimating the galaxy redshifts from the \lya\ break alone without an additional DLA component, we also model each spectra with only the IGM component where $z_{\rm break}$ and $x_{\rm HI}$ are left as free parameters. This scenario represents the cases where the exact systemic redshifts cannot be determined from any identifiable emission lines, and only serves as a way to quantify the \lya-break redshift bias.  
    \item {\bf ISM only.} In this iteration, we remain agnostic about the exact location of the bulk \hi\ gas. This is due to the expected degeneracies between the IGM component contributing to $x_{\rm HI}$ and the dense, local gas contributing to $N_{\rm HI}$. We thus simply model the \lya\ damping wing as a DLA with the $N_{\rm HI}$ output representing the total amount of \hi\ gas integrated along the line of sight to each source. This provides an upper bound on the local \hi\ gas density in each galaxy. 
\end{itemize}

\begin{figure}[!t]
    \centering
    \includegraphics[width=9.2cm]{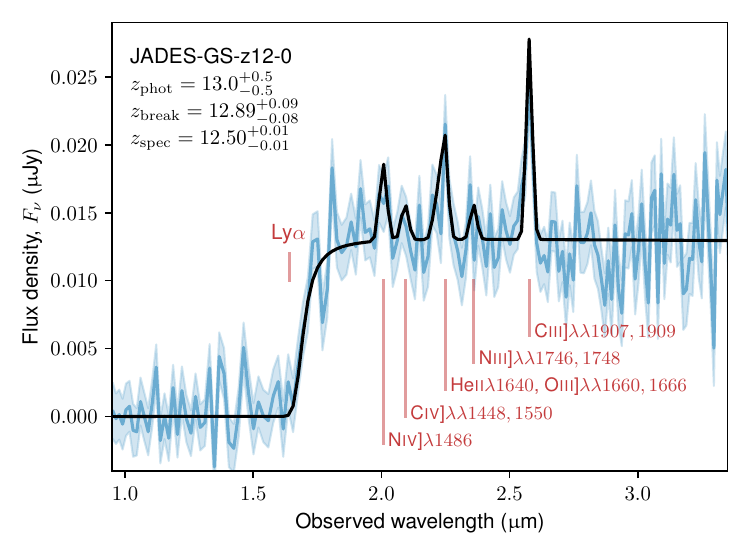}
    \caption{Combined DLA and nebular emission-line model fit (black) to an example NIRSpec-Prism spectrum (blue). The spectrum was first presented in \citet{CurtisLake23} and then \citet{DEugenio24}, but here shown is the DJA-v4.5 version which is a combination of the two. The blue solid line and the shaded region show the flux and error spectrum, respectively. The \lya\ redshift and the most prominent rest-frame UV nebular emission lines are marked at the best-fit spectroscopic redshift $z=12.50^{+0.01}_{-0.01}$ with a derived \hi\ column density of $\log (N_{\rm HI}/{\rm cm^{-2}}) = 22.36^{+0.16}_{-0.19}$. The photometric redshift $z_{\rm phot} = 13.0^{+0.5}_{-0.5}$ is from \citet{Robertson23}. The \lya-break redshift $z_{\rm break}=12.89^{+0.09}_{-0.08}$ is derived with a fixed IGM neutral hydrogen fraction of $x_{\rm HI}=1$ without including a DLA component. }
    \label{fig:dlamodel}
\end{figure}

The general \lya\ damping wing modeling is described in detail in \citet{Pollock26}. Briefly, we use {\tt dynesty} \citep{Speagle20} to perform the statistical modeling for each of the iterations described above to estimate the posterior distributions for each of the free parameters and the total evidence used for model comparison. We consider a broad range in local \hi\ column density, $\log (N_{\rm HI}/{\rm cm^{-2}})\in[18,24]$, and a zero to fully neutral IGM, $x_{\rm HI}\in[0,1]$. Due to the low spectral resolution of the NIRSpec Prism observations, there is a large degeneracy between these two quantities at lower $N_{\rm HI} \lesssim 10^{21}\,{\rm cm}^{-2}$ \citep{Heintz24_DLA,Huberty25,Pollock26}, such that only galaxies with strong DLAs can be robustly attributed to the local ISM or CGM \citep{Heintz25}. The rest-frame UV galaxy continuum is modeled as a simple power-law, with $F_\lambda \propto \lambda^{\beta_{\rm UV}}$ with $\beta_{\rm UV}\in[-4,0]$, superimposed with Gaussian emission-line profiles at $z_{\rm spec}$. An example spectrum and the best-fit models from the two scenarios are shown in Fig.~\ref{fig:dlamodel}, highlighting the redshift bias from considering the \lya\ break and IGM component alone.     

%\subsection{Properties from the spectral energy distributions}
\subsection{Physical properties and star-formation histories}

% Describe briefly the Bagpipes SED outputs. Focus on SFRs. 
%Add also here sizes.

For each of the compiled galaxies at $z>10$, we also derive their physical properties. First, we perform spectro-photometric fitting of their spectral energy distributions (SEDs) with the {\tt Bagpipes} tool \citep{Carnall2018}, as described in detail in \citet{Pollock26}. Briefly, we photometrically correct the spectra by scaling the overall flux density to match the magnitude in the closest rest-frame UV filter (typically \jwst/NIRCam's F200W) and for differential slit loss based on the location of the source on the shutter. The redshifts of each galaxy is fixed to that derived from the most prominent emission lines or from the constraints from the \lya\ damping wings. We adopt a non-parametric continuity star-formation history \citep{Leja19}, with lookback time bins in 10 equally spaced, logarithmic time intervals, and a flexible dust attenuation curve \citep[][modified from \citealt{Calzetti2000}]{Salim18} to model the photometrically-scaled spectra and the available photometry simultaneously (see Xiao et al., submitted and Gottumukkala et al. in prep. for further details on the {\tt Bagpipes} modeling done for the entire DJA spectroscopic repository, including constraints from the continuum and emission-line properties). From the output SEDs, we mostly consider the total stellar masses and the star-formation rates (SFRs) integrated over two time intervals, 10\,Myr and 100\,Myr. The ratios of these, in particular, have been used to quantify the `burstiness' of the galaxies in the recent literature \citep[e.g.,][]{Endsley24,RobertsBorsani24,Simmonds24,Simmonds25,Kokorev25}. The SFR estimates over the past 100\,Myr timescales is mainly represented by the overall UV luminosity of the galaxies. In contrast, the SFRs derived on 10\,Myr timescales represent the instantaneous or current burst of the most massive ($M_\star \gtrsim 10\,M_\odot$) star formation, as also typically traced by the Balmer recombination lines \citep[e.g.,][]{Kennicutt98}. Due to the high redshift of the sample, the strongest of these (H$\beta$ and H$\alpha$) are unfortunately redshifted out of the \jwst/NIRSpec Prism wavelength coverage. 

\begin{figure*}[!t]
    \centering
    \includegraphics[width=17cm]{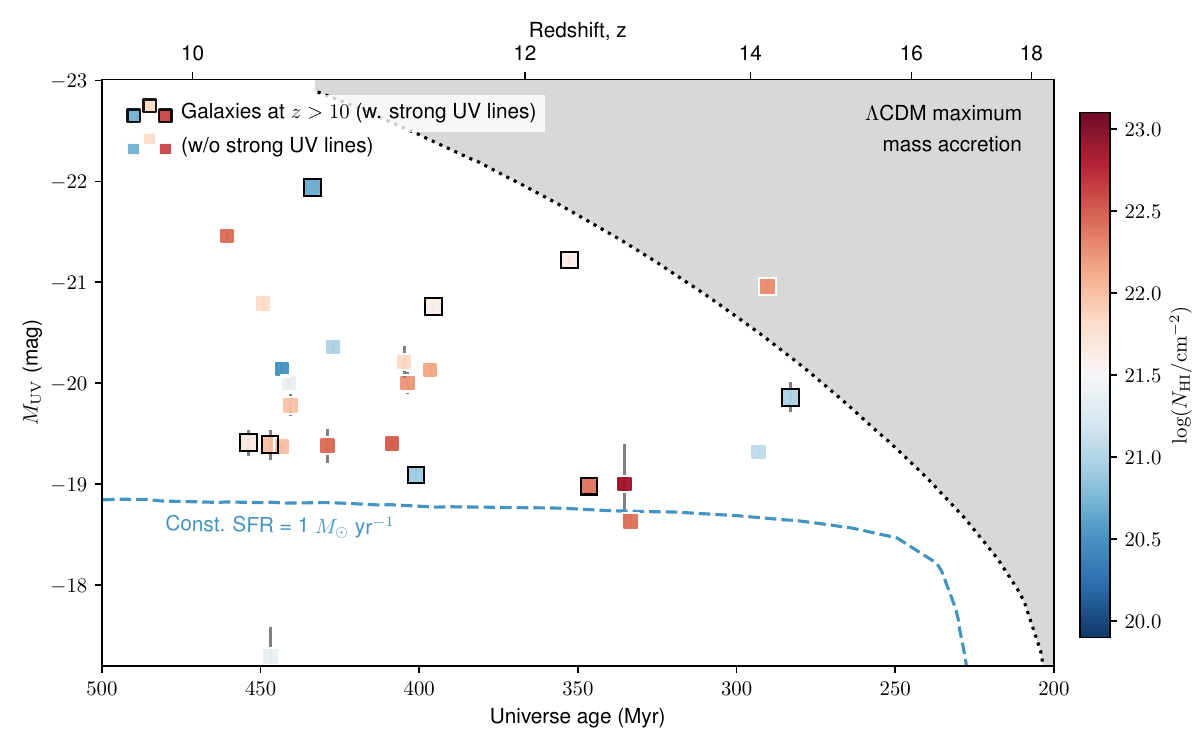}
    \caption{Absolute UV magnitude, $M_{\rm UV}$, as a function of cosmic time for the full $z>10$ galaxy sample. Each galaxy is color-coded according to their derived \hi\ gas column density, $N_{\rm HI}$. The sources with strong rest-frame UV lines detected in their spectra are highlighted with black edge-colors. For non-constrained DLA fits, the total \hi\ column density assuming an ISM origin is denoted by the color (typically those with $\lesssim 10^{21.5}\,{\rm cm}^{-2}$). For reference are shown tracks of constant star formation at $1\,M_\odot {\rm yr}^{-1}$ (blue) and the average maximum possible accretion scenario allowed in standard $\Lambda$CDM cosmology \citep[gray region; see][]{Dekel13}, both adapted from \citet{Kokorev24}.}
    \label{fig:muvz}
\end{figure*}

Intriguingly, the \ciii]\,$\lambda 1909$ transition might be a robust and more feasible alternative tracer of the instantaneous SFR at $z>10$. This feature is typically among the strongest and most prominently detected emission lines in the rest-frame UV at high redshifts \citep{DEugenio24,Cunningham24}, providing first a reliable anchor to estimate the systemic redshift of the sources. This especially so since \lya\ will be mostly suppressed by the large neutral hydrogen fraction in the IGM and any potential local \hi\ gas at this epoch. Then, given that the upper state of the \ciii]\,$\lambda 1909$ transition is mainly populated by photo-ionization of massive stars in sub-solar ($Z\lesssim 0.5\,Z_\odot$) systems with young stellar populations, the total emission should thus trace star formation on $\sim 3-10$\,Myr timescales \citep[e.g.,][]{Erb10,Stark14,Stark15}. Indeed, we find evidence of the \ciii]\,$\lambda 1909$ line luminosities being correlated with the SFRs derived from the SED on 10\,Myr timescales and from H$\beta$ in a lower-redshift sample ($z=9-10$, from \citealt{Pollock25}), see Appendix~\ref{sec:appA}, with a best-fit log-linear relation $\log(L_{\rm CIII]} / {\rm erg\,s^{-1}}) = a\times \log ({\rm SFR}_{\rm 10\,Myr}/M_\odot\,{\rm yr}^{-1}) + b$, with slope and intercept $a=0.69\pm 0.14$ and $b=41.46\pm 0.14$, respectively. This indicates that the \ciii]\,$\lambda 1909$ line transition can indeed be used to represent and predict the instantaneous SFR of galaxies at $z>10$, though with the caveat that particularly metal-poor or very high-redshift systems may be carbon-enhanced when dominated by Pop\,III stellar yields \citep{DEugenio24,Nakajima26,Pollock26b}. Further, this can be compared to the UV luminosity of the spectra to gauge the burstiness of the galaxies directly without relying on modeling their SED and the assumptions entering into this. 
% Cunningham+24: suggesting that UV-bright galaxies with strong Ly α and C III] emission may be undergoing a more ‘bursty’ phase of star formation, aiding the efficient clearing of channels through which Ly α can escape, consistent with the ‘expanding shell model’. 

The main physical properties of each target derived in this work is summarized in Table~\ref{tab:tab2}. In the following analysis, we will mostly rely on the SFRs derived from the SED modeling to be consistent with the literature and for more direct comparison for the cases where \ciii]\,$\lambda 1909$ is not detected. Here we also summarize the main physical properties of the target galaxies, in particular related to their local dense \hi\ gas reservoirs.

%%% TABLE %%%%
\begin{table*}
\begin{center}    % used for centering table
\caption{Summary of the main physical properties studied here for the compiled galaxies at $z>10$.} % title of Table
\label{tab:tab2}      % is used to refer this table in the text
\setlength\tabcolsep{0.37cm}
\renewcommand{\arraystretch}{1.2}
\begin{tabular}{c c c c c c }
\hline\vspace{0.1cm}%\hline                        % inserts double horizontal lines
Source ID & $z_{\rm spec}$ & $\log \Sigma_{\rm SFR}$  & $\log N_{\rm HI}$ & $F_{\rm [CIII]}$ & ${\rm SFR_{10\,Myr}} / {\rm SFR_{100\,Myr}}$ \\ % table heading
&&  ($M_\odot$\,yr$^{-1}$\,kpc$^{-2}$) & (cm$^{-2}$) & ($\times 10^{-19}$\,erg\,s$^{-1}$\,cm$^{-2}$)  \\
% (deg) & (deg) & & & & \\
 (1) &  (2) &  (3) &  (4) & (5) & (6) \\
\hline   % inserts single 
MoM-z14         & 14.4441 & $2.16\pm 0.08$ & $< 21.0$ & $2.41^{+0.77}_{-0.76}$ & $1.20^{+0.47}_{-0.36}$ \\
JADES-GS-z14-0  & 14.1832 & $1.50\pm 0.03$ & $22.25^{+0.11}_{-0.13}$ & $2.07^{+0.65}_{-0.60}$ &  $1.17^{+0.33}_{-0.32}$\\
JADES-GS-z14-1  & 14.0848 & $1.27\pm 0.01$ & $< 21.1$ & $< 1.3$ &  $1.48^{+0.40}_{-0.32}$ \\
JADES-GS-z13-0  & 12.8479 & $1.81\pm 0.03$ & $22.38^{+0.14}_{-0.16}$ & $1.19^{+0.36}_{-0.38}$ &  $1.13^{+0.41}_{-0.33}$ \\
JADES-GS-z13-1-LA & 12.7959 & $2.44\pm 0.01$ & $22.86^{+0.16}_{-0.22} $ & $< 0.5$ &  $--$ \\
JADES-GS-z12-0  & 12.4978 & $0.49\pm 0.30$ & $22.36^{+0.18}_{-0.19} $ & $1.66^{+0.24}_{-0.24} $ &  $4.13^{+0.43}_{-0.40}$ \\
GHZ2/GLASS-z12  & 12.3420 & $3.34\pm 0.10$ & $21.60^{+0.52}_{-0.21} $ & $8.07^{+1.00}_{-1.01} $ &  $3.77^{+0.44}_{-0.32}$ \\
CAPERS-88619    & 11.3637 & $0.85\pm 0.06$ & $< 21.6$ & $5.26^{+0.92}_{-0.93}$ &  $1.52^{+0.43}_{-0.37}$ \\
CEERS-16943     & 11.3408 & $0.91\pm 0.04$ & $22.12^{+0.24}_{-0.52}$ & $5.40^{+1.06}_{-1.13}$ &  $1.67^{+0.35}_{-0.30}$ \\
JADES-GS-z11-1  & 11.2499 & $1.80\pm 0.06$ & $< 20.9$ & $2.93^{+0.84}_{-0.81}$ &  $2.46^{+0.35}_{-0.26}$ \\
CEERS-10        & 11.1962 & $--$           & $22.21^{+0.15}_{-0.32}$ & $< 3.6$ &  $1.36^{+0.35}_{-0.32}$ \\
CAPERS-126973   & 11.1736 & $0.60\pm 0.30$ & $< 21.8$ & $< 5.3$ &  $1.25^{+0.41}_{-0.34}$ \\
JADES-GS-z11-0  & 11.0979 & $0.50\pm 0.30$ & $22.47^{+0.24}_{-0.17}$ & $< 1.9$ &  $0.84^{+0.36}_{-0.27}$ \\
CAPERS-22637    & 10.7457 & $0.51\pm 0.30$ & $< 21.0$ & $< 4.3$ &  $1.47^{+0.38}_{-0.28}$ \\
MoM-z11         & 10.7123 & $0.41\pm 0.30$ & $22.41^{+0.24}_{-0.32}$ & $< 5.8$ &  $1.31^{+0.42}_{-0.29}$ \\
GNz11           & 10.6296 & $3.04\pm 0.12$ & $< 20.7$ & $15.3^{+1.03}_{-1.09} $ &  $4.40^{+0.28}_{-0.23}$ \\
GS-z10          & 10.5052 & $0.12\pm 0.01$ & $< 22.0$ & $< 4.7$ &  $2.04^{+0.36}_{-0.33}$ \\
CAPERS-136645   & 10.4980 & $0.38\pm 0.30$ & $21.38^{+0.37}_{-0.55}$ & $< 6.7$ &  $0.53^{+0.50}_{-0.44}$ \\
JADES-20176151  & 10.4599 & $0.51\pm 0.30$ & $21.98^{+0.18}_{-0.20}$ & $< 1.9$ &  $2.00^{+0.32}_{-0.24}$ \\
Abell-2744-22302& 10.4597 & $0.75\pm 0.35$ & $< 20.5$ & $6.62^{+1.82}_{-1.76} $ &  $0.15^{+1.09}_{-0.86}$ \\
JADES-GS-z10-0  & 10.3989 & $-0.79\pm 0.30$ & $< 21.4$ & $< 1.8$ &  $1.47^{+0.37}_{-0.28}$ \\
Abell-2744-23984& 10.3591 & $1.06\pm 0.35$ & $< 21.8$ & $< 6.2$ &  $2.85^{+0.38}_{-0.29}$ \\
CAPERS-82699    & 10.3960 & $0.75\pm 0.07$ & $< 22.0$ & $6.54^{+2.13}_{-2.27}$ &  $--$ \\
CAPERS-109917   & 10.2807 & $0.50\pm 0.06$ & $< 21.7$ & $5.17^{+1.24}_{-1.21}$ &  $2.26^{+0.56}_{-0.40}$ \\
MACS0647-3349   & 10.1716 & $2.76\pm 0.13$ & $22.40^{+0.07}_{-0.07}$ & $16.6^{+2.71}_{-2.83}$ & $1.23^{+0.23}_{-0.22}$ \\
\hline          %inserts single line
\end{tabular} 
\end{center}  
\textbf{Notes.} Column (1): Source ID. Column (2): Spectroscopic redshift from the identified emission-lines. Column (3): SFR surface density. Column (4): \hi\ column density, where upper limits denote the maximum $N_{\rm HI}$ in the IGM and ISM along the line of sight. Column (5): \ciii]\,$\lambda 1909$ line flux. Upper limits are quoted at $3\sigma$ significance. Column (6): Burstiness parameter, here quantified as ${\rm SFR_{10\,Myr}} / {\rm SFR_{100\,Myr}}$ from the SED model. All relevant quantities have been scaled to the match the photometry and corrected for lensing magnification. 
%{\bf References.} ??
\end{table*}
%%%%%%%%%%%%%%

\section{Results}
\label{sec:res}

\subsection{Rest-frame UV brightness}

To investigate the properties of the galaxy population at $z>10$, %within the first 500\,Myr of cosmic time, we 
we first consider their absolute UV magnitude as a function of cosmic time and derived \hi\ column density in Fig.~\ref{fig:muvz}. Overall, the compiled sample galaxies span a range of luminosities from $M_{\rm UV} = -18$ to $-22$\,mag and \hi\ column densities $N_{\rm HI} = 10^{20.5}-10^{22.5}\,{\rm cm}^{-2}$. The \hi\ column densities here and in the figure are assuming the maximum allowed ISM contribution for the non-converged cases. We contextualize these results by comparing to the evolutionary track of a constant star-formation rate of ${\rm SFR} = 1\,M_\odot {\rm yr}^{-1}$ and predictions for the UV luminosity of a galaxy following a maximal possible accretion scenario allowed in standard $\Lambda$CDM cosmology assuming a smooth gas accretion to SFR \citep[][]{Dekel13}, see also \citet{Kokorev24}. The large majority of the sample galaxies show higher equivalent $M_{\rm UV}$ than the constant SFR track, consistent with the typical galaxy at $z\sim 10$ having higher SFRs than one solar mass per year, at least for stellar masses $M_\star \gtrsim 10^{8}\,M_\odot$ \citep[e.g.,][]{Heintz23_FMR,Pollock25}. The two galaxies with fainter $M_{\rm UV}$ are JADES-GS-z13-0 at $z=12.85$ and JADES-GS-z10-0, respectively \citep{CurtisLake23}, consistent with their low stellar masses and SFRs derived from photometry \citep{Robertson23}.  

On the opposite end, only the source JADES-GS-z14-0 at $z=14.18$ \citep{Carniani24a} appears to be located in the $\Lambda$CDM-excluded region. This is at odds with the expected scenario where this source would be observed in a snapshot during a particular bright or `bursty' evolutionary phase \citep{Mason23,Munoz26}, since it shows no apparent strong emission lines indicating recent starburst activity and has a relative low ${\rm SFR}_{\rm 10\rm Myr} / {\rm SFR_{100\,Myr}} \approx 1.2$. Intriguingly, this galaxy is embedded in a pristine, dense \hi\ gas reservoir, with a derived \hi\ column density $N_{\rm HI}\gtrsim 10^{22}$\,cm$^{-2}$ \citep[see also][]{Carniani25,Heintz25_z14}, pointing to this source being a young, nascent galaxy. The three other galaxies occupying the bright end of the distribution and straddling the excluded region are MoM-z14 at $z=14.44$ \citep{Naidu25}, GHZ2/GLASS-z12 at $z=12.34$ \citep{Castellano24}, and GN-z11 at $z=10.63$ \citep{Oesch16,Bunker23_gnz11}. Notably, these three galaxies all have prominent rest-frame UV emission lines detected in their spectra and neither show evidence for strong DLAs hinting at bulk \hi\ gas in the ISM, with \lya\ transmission profiles consistent with that expected for a partly to fully neutral hydrogen fraction in the IGM \citep[see also][]{Pollock26}. This is in stark contrast to JADES-GS-z14-0 with almost no rest-frame UV lines and a prominent DLA, potentially suggesting two different physical mechanisms powering the extreme UV brightness of these sources or a large variety in the metal production and ionizing capabilities in the the line of sight to these sources at similar early epochs.  

% Note on general population statistics. 
For the general galaxy population at $z>10$, we do not find any significant evidence for the UV-brightest galaxies showing the strongest rest-frame UV emission lines nor high column density DLAs. Indeed, the former is seen in both the brightest and one of the faintest galaxies in our sample. Similarly, both low and high \hi\ gas column density sightlines are observed in the faintest and brightest galaxies. This hints at a more subtle effect driving the UV brightness or efficiency of the star formation in these early, nascent galaxies. 

\subsection{Bursty star formation}

\begin{figure}[!t]
    \centering
    \includegraphics[width=9.5cm]{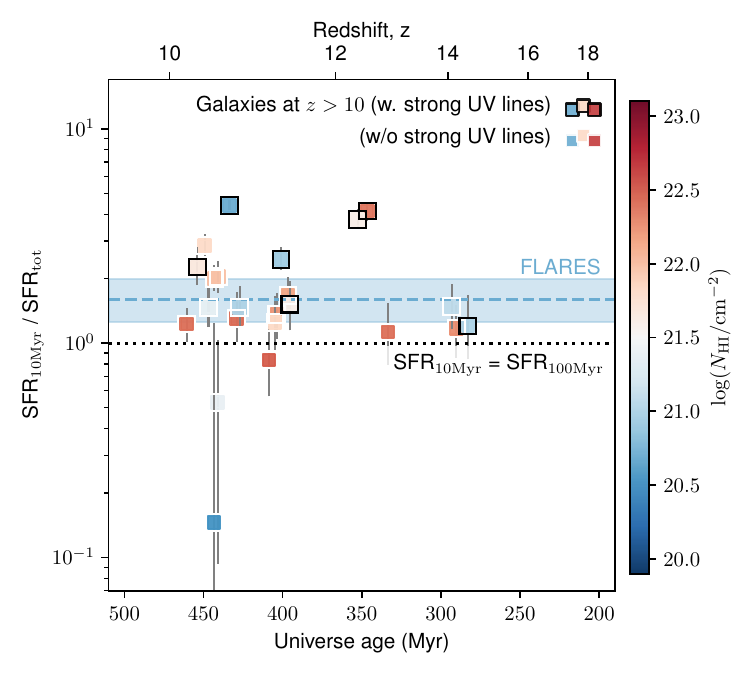}
    \caption{Fraction of short- ($10$\,Myr) to long-lived ($\sim 100$\,Myr) star formation as a function of cosmic time. The symbol notation follows Fig.~\ref{fig:muvz}. For comparison is shown the model predictions from the FLARES simulations \citep{Wilkins23b}. The dotted line shows the unity ratio for a constant star-formation rate over the past $\sim 100$\,Myr. }
    \label{fig:sfrburst}
\end{figure}

We then consider the effect of dense \hi\ gas on the `burstiness' of the galaxies at $z>10$, here quantified as the fraction of short- ($10$\,Myr) to long-lived ($\sim 100$\,Myr) star-formation rate, ${\rm SFR_{10\,Myr}}/{\rm SFR_{100\,Myr}}$ \citep[as is typically done in the literature, e.g.,][]{Endsley25,Cole25,Kokorev25,CarvajalBohorquez25,Simmonds25}, integrated over the star-formation history from the SED modeling. In Fig.~\ref{fig:sfrburst}, we show ${\rm SFR_{10\,Myr}}/{\rm SFR_{100\,Myr}}$ as a function of cosmic time, again color-coded with $N_{\rm HI}$. Now a more clear picture emerges: Five out of seven galaxies with strong rest-frame UV lines are observed in a bursting phase with a recent SFR $2-5\times$ higher than the average over the past 100\, Myr. This is consistent with the scenario where these are the most actively star forming systems with intense ISM radiation fields driving the high-ionization emission lines. Indeed, the strongest \ciii]\,$\lambda 1909$ emitters are also observed with the highest star-formation rate surface densities \citep[][see also Sect.~\ref{ssec:ks} below]{RobertsBorsani25}. The one exception is MoM-z14 which show prominent rest-UV lines but without significant evidence for an excess in ${\rm SFR_{10\,Myr}}/{\rm SFR_{100\,Myr}}$ from the {\tt Bagpipes} modeling. We note, however, that \citet{Naidu25} found evidence for a rise in SFR on shorter timescales when modeling the SED with {\tt Prospector} \citep{Leja17,Johnson21}, potentially up to an order of magnitude within $\approx 10$\,Myr which places MoM-z14 at equally (or even higher) burstiness compared to the other galaxies at $z>10$ with prominent rest-UV lines. The relative \ciii]\,$\lambda 1909$ line to UV continuum luminosity is close to the median of the full sample, so the actual burstiness of MoM-z14 remains inconclusive. 

Overall, we find that full sample shows evidence for an excess of short- to long-lived SFR, with a mean and median value of ${\rm SFR_{10\,Myr}}/{\rm SFR_{100\,Myr}} = 1.74$ and $1.31$, respectively, with a scatter of 0.3\,dex. This is consistent with predictions from the hydrodynamical First Light And Reionization Epoch Simulations \citep[FLARES;][]{Lovell21,Wilkins23a,Wilkins23b}, which is mainly a representation of the smooth dark matter accretion rate and the active assembly and build-up of stars. Only galaxies above this relation should be treated as being in a strong burst phase. We find no apparent trend between the \hi\ column density and ${\rm SFR_{10\,Myr}}/{\rm SFR_{100\,Myr}}$, with a large variation in column densities ($10^{21}-10^{22.5}$\,cm$^{-2}$) for the most bursty galaxies. However, we note that the majority of the sample with prominent UV lines are found with lower $N_{\rm HI} < 10^{22}\,{\rm cm}^{-2}$, and vice versa, all galaxies with $N_{\rm HI} > 10^{22}\,{\rm cm}^{-2}$ show no evidence for strong rest-UV lines except for JADES-GS-z12-0 \citep{CurtisLake23,DEugenio24}. % Interestingly, the source UNCOVER-37126 at $z=9.85$ with the highest ${\rm SFR_{10\,Myr}}/{\rm SFR_{tot}}$ also shows one of the strongest DLAs with $\log (N{\rm_{HI}/cm^{-2}}) = 22.53^{+0.09}_{-0.09}$.   excluded due to low redshift

Generally, we note that the short- to long-lived SFR ratio used as a proxy for the burstiness of the target galaxies here, likely also or alternatively reflects the young ages of the average stellar population in these systems. Examining the star-formation histories (SFHs) indeed shows that the majority of galaxies in the sample have predicted formation times only $\sim 30-50$\,Myr before the time of observation. The excess ${\rm SFR_{10\,Myr}}/{\rm SFR_{100\,Myr}}$ ratio thus mainly trace the current rising SFHs of the target, UV-bright galaxies. Complementarily, \citet{Endsley24} found evidence for the brightest galaxies at $z\gtrsim 6$ tend to have more very-recently rising SFHs compared to fainter galaxies at the same redshift, albeit with a large scatter. The most bursty galaxies were also found to have the most prominent nebular emission lines \citep{Endsley24,RobertsBorsani25}, in agreement with our results.

\subsection{Star-formation efficiency} \label{ssec:ks}

\begin{figure}[!t]
    \centering
    \includegraphics[width=9.3cm]{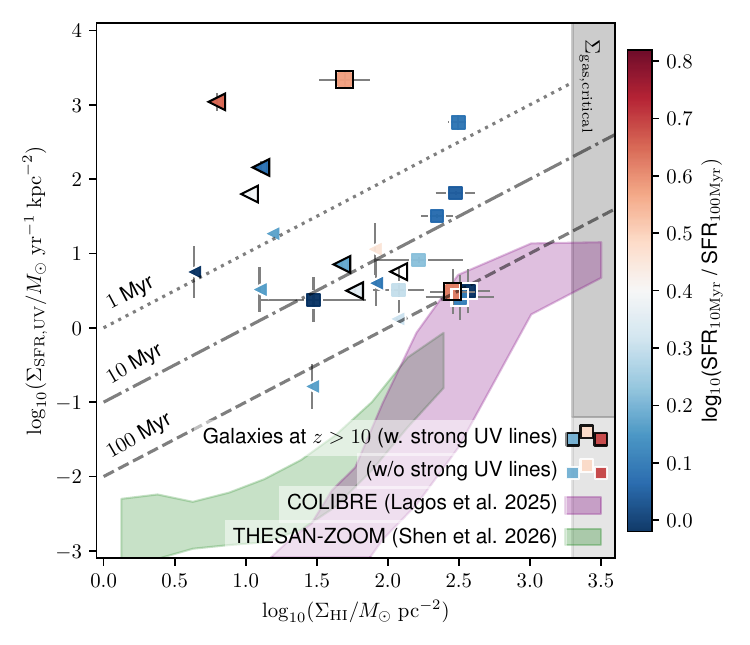}
    \caption{UV-derived star-formation rate surface density, $\Sigma_{\rm SFR,UV}$, as a function of the \hi\ gas mass density, $\Sigma_{\rm HI}$. The symbol notation follows Fig.~\ref{fig:muvz}, but here the galaxies are color-coded according to their `burstiness' parameter, ${\rm SFR_{10\,Myr}} / {\rm SFR_{100\,Myr}}$. For each galaxy, $\Sigma_{\rm HI}$ is a direct unit-conversion from $N_{\rm HI}$ where the well-converged cases are shown by the squares and the total \hi\ gas upper bounds are shown as the leftward facing triangles for the unconstrained sources. The diagonal lines represent a range of equivalent gas depletion timescales. For comparison, model predictions from the {\tt COLIBRE} \citep{Lagos25} and {\tt THESAN-ZOOM} \citep{Kannan25,Shen26} simulations are shown as the purple and green shaded regions. The critical gas surface density, $\Sigma_{\rm gas,critical}\sim 2000\,M_\odot \,{\rm pc}^{-2}$, where stellar feedback becomes ineffective at $z\sim 10$ is marked as well \citep{Somerville25}.}
    \label{fig:ksplot}
\end{figure}

To investigate more directly the star-formation efficiency (SFE) of the target galaxies at $z>10$ based on the availability of the dense, local gas, we compare the SFR surface density, $\Sigma_{\rm SFR}$, to the \hi\ gas surface density, $\Sigma_{\rm HI}$, as a function of ${\rm SFR_{10\,Myr}}/{\rm SFR_{100\,Myr}}$ in Fig.~\ref{fig:ksplot}. In this formalism, we quantify the SFE as the offset from the canonical Kennicutt-Schmidt relation. We determine $\Sigma_{\rm SFR}$ as ${\rm SFR_{UV}} / (2\pi R^2_{e})$ where ${\rm SFR_{UV}}$ is the star-formation rate derived from the UV continuum and $R_e$ is the effective half-light radius measured in the closest available rest-frame UV filter (typically F150W or F200W at $z=10-14$), either adopted from the DJA photometric catalog or the available literature for select sources. This effectively provides a lower bound on the actual SFR surface density, given the typical bursty nature of the galaxies. The \hi\ gas surface density is derived directly from the line-of-sight \hi\ column density. This was first established by \citet{Pollock26}, arguing that the integrated \hi\ column density over the entire UV-emitting region covered by the \jwst\ shutter represents the total-average gas surface density (though see their work for more details and caveats). We also include predictions for the same quantities from the {\tt COLIBRE} \citep{Schaye26,Lagos25} and {\tt THESAN-ZOOM} \citep{Kannan25,Shen26} hydrodynamical simulations.

We observe that all but one of the galaxies with the most prominent rest-frame UV nebular emission lines are among the population with the highest $\Sigma_{\rm SFR}$ and shortest gas depletion times, defined as $t_{\rm dep} \equiv  \Sigma_{\rm gas}/\Sigma_{\rm SFR} \lesssim 10$\,Myr \citep{Kennicutt98}. This is related to the star-formation efficiency via ${\rm SFE} = t_{\rm ff}/t_{\rm dep}$, where $t_{\rm ff}$ is the free-fall time of the gas cloud, found to be 25\,Myr on average for galaxies at $z>9$ \citep{Pollock26}. These results are consistent with a scenario where these particular galaxies are caught during a recent burst or rising star formation episode \citep[see e.g. also][]{RobertsBorsani25}. While the {\tt COLIBRE} and {\tt THESAN-ZOOM} simulations generally predict a similar range in \hi\ gas surface densities at similar redshifts and stellar masses, the typical SFR surface densities and thereby predicted star-formation efficiencies are substantially lower than derived here. This suggests that these simulations generally underestimate the actual star-formation efficiency of these early galaxies, and less so related to the short-lived, bursty star formation episodes. We note that most of the sample galaxies have gas surface densities below the critical limit, $\Sigma_{\rm gas,critical} \sim 2000\,M_\odot\,{\rm pc}^{-2}$ where stellar feedback begins to become ineffective. This is somewhat at odds with predictions from galaxy formation models, that suggest that the majority of galaxies at $z>10$ should enter this regime \citep{Somerville25}, both for low (high) gas fractions of $f_{\rm gas} = M_{\rm gas}/(f_{\rm b} M_{\rm h}) = 0.2$ (1.0). Finally, we highlight that the galaxies at $z>10$ show a continuous distribution of $\Sigma_{\rm SFR}$, in contrast to the recently proposed dichotomy in effective galaxy sizes at the same redshifts \citep{Harikane25,Naidu25}. 

We note that the most intense star-forming systems approach the classical Heckman threshold for starburst galaxies at the theoretical limit of $\Sigma_{\rm SFR}\sim 100\,M_\odot\,{\rm yr}^{-1}\,{\rm kpc}^{-2}$, above which feedback-driven outflows are expected to become ubiquitous. This could potentially explain their lower inferred gas surface densities or indicate that the star formation is strongly driven by the external processes such as rapid gas infall that enables continuous gas replenishment.  

Comparing the SFE or gas depletion timescales of the sample galaxies to their derived burstiness, we also observe a continuous distribution where both bursty and less active star-forming galaxies appear to span the entire parameter space from the most efficient star-formation (or shortest gas depletion times), to long $\sim 100$\,Myr depletion timescales and relatively inefficient star formation. In particular, some of the most bursty galaxies with ${\rm SFR_{10\,Myr}} / {\rm SFR_{100\,Myr}} > 3$ show \hi\ gas surface densities from $\Sigma_{\rm gas} \sim 10$ to $\sim 300\,M_\odot\,{\rm pc}^{-2}$. The galaxies with ${\rm SFR_{10\,Myr}}\sim {\rm SFR_{100\,Myr}}$ are observed to span a similar range. This suggests that we are likely observing an evolving population of galaxies with rapidly rising SFHs that has not yet exhausted their available gas reservoirs to systems that have consumed or ejected most of their available gas in this scenario, showing an artificially high SFE. The latter will thus not be able to sustain their intense star formation unless there is substantial gas replenishment from the intergalactic medium. Intriguingly, \citet[][see also \citealt{Simmonds25}]{Munoz26} find typical star-formation burst timescales of approximately 20 Myr, which is consistent with the gas free-fall times derived by \citet{Pollock26} at $z>9$ and with expectations from supernova feedback \citep{Furlanetto22}. Complementary to this, \citet{Langeroodi24} find an apparent dearth of bursting galaxies at $z\sim 4$, coincident with the redshift where most of the available IGM \hi\ gas is expected to have accreted onto galaxies \citep[e.g.,][]{Heintz22}. 

We argue that all these results point to a physical scenario where excessive \hi\ gas infall at $z>10$ drive the rapid and stochastic star formation, manifesting as an observed large variation in $\Sigma_{\rm gas}$ vs $\Sigma_{\rm SFR}$ due to the approximately equal star forming burst and gas depletion timescales at these redshifts. This scenario was first proposed by \citet{Tacchella20} based on simple stochastic star formation models, and later also reproduced in the {\tt THESAN-ZOOM} simulations \citep{McClymont25}.
% (https://ui.adsabs.harvard.edu/abs/2020MNRAS.497..698T) have already highlighted from simple stochastic SF models that at high-z inflow variability is crucial... and this has also been seen in THESAN-ZOOM, where burstiness is driven by both intneral stellar feedback and external inflows; the latter is more important at z>8 (https://ui.adsabs.harvard.edu/abs/2025MNRAS.544..513M)

\section{The impact of strong Ly$\alpha$ absorption on redshift estimates and the UV LF} \label{sec:uvlf}

One of the early surprise discoveries with \jwst\ was the slight systematic overpredictions of photometric \citep[e.g.,][]{Finkelstein24,Hainline24b,Willott24} and Ly$\alpha$ break \citep[e.g.,][]{CurtisLake23,ArrabalHaro23a,Fujimoto23_ceers} redshifts when compared to the first systemic nebular emission-line measurements. This discrepancy was shown to originate mainly from the high prevalence of strong DLAs in the spectra of a large fraction galaxies at $z\gtrsim 9$ \citep{Heintz24_DLA}. The integrated line-of-sight \hi\ column densities of some of these galaxies are so high that the photometric redshifts are overestimated by $\Delta z = 0.5-1$ \citep{Serjeant23,Hainline24b,Witstok25_O3,Donnan26}. Later studies on larger statistical samples \citep{Asada25,Heintz25,Hainline26} show that photometrically inferred redshifts are typically overestimated by $\sim 0.2$ on average at $z>5$. In this section, we investigate the redshift bias particular for the $z>10$ sample galaxies, with the main goal to determine the impact these overestimates have on the high intrinsic UV brightness inferred for these galaxies.   

\begin{figure}[!t]
    \centering
    \includegraphics[width=9cm]{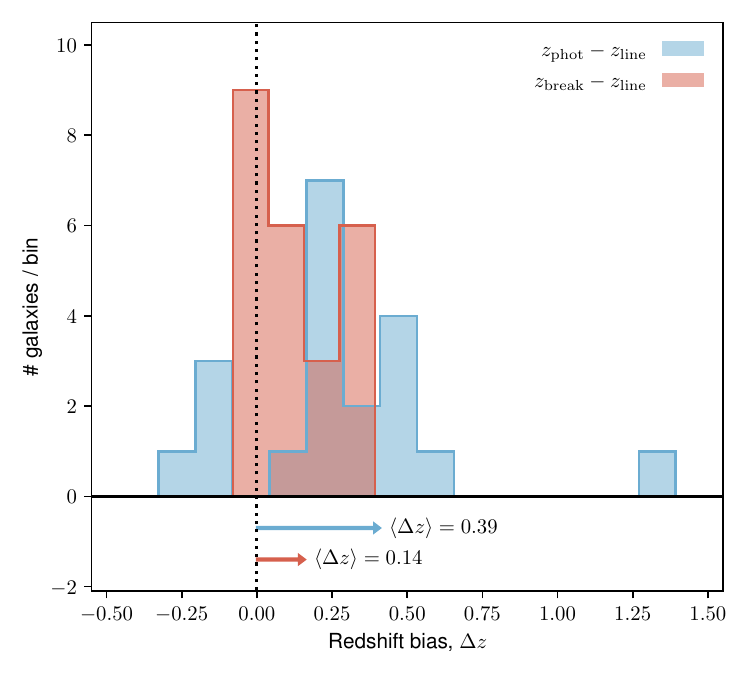}
    \caption{Histogram of the redshift-bias distribution for the galaxies at $z>10$. The redshift offsets represent $z_{\rm phot}-z_{\rm line}$ (blue) and $z_{\rm break}-z_{\rm line}$ (red). The mean redshift biases for the two distributions are shown by the bottom blue and red arrows, respectively. }
    \label{fig:zbiasdist}
\end{figure}

First, we show in Fig.~\ref{fig:zbiasdist} the `redshift bias', $\Delta z = z_{\rm break}-z_{\rm line}$, distribution, comparing the photometric $z_{\rm phot}$ and \lya\ break $z_{\rm break}$ redshifts to the spectroscopically derived line redshifts, $z_{\rm spec}$. The photometric redshifts are either adopted from the literature for single galaxy measurements or from the photometric DJA catalogs \citep{Valentino23} based on {\tt Eazy} \citep{Brammer_eazy}, as summarized in Table~\ref{tab:tab1}. The \lya\ break redshifts are measured conservatively by assuming a fully neutral hydrogen fraction in the IGM, $x_{\rm HI}$, for the \lya\ damping wing modeling leaving only the redshift as a free parameter, as detailed in Sect.~\ref{ssec:dla}. For a partly ionized IGM the redshift bias can potentially be higher. Leaving $x_{\rm HI}$ free in the modeling, generally has a negligible effect on the inferred redshifts and in most cases only show a flat posterior with a modest degeneracy between $z_{\rm break}$ and $x_{\rm HI}$. As expected, we find that the galaxies without a strong DLA in their spectra generally have $z_{\rm phot}$ and $z_{\rm break}$ consistent with $z_{\rm spec}$. However, there is a clear systematic trend for a prominent redshift bias in both quantities, with mean offsets of $\Delta z_{\rm phot} = 0.39$ and $\Delta z_{\rm break} = 0.14$, respectively, with a clear higher-offset tail, particular for $\Delta z_{\rm phot}$. 
%Show PDF(z) of photo-z, Lya break, and lines for an example galaxy? 

%%% TABLE %%%%
\begin{table}[!t]
\caption{Spectroscopic UV luminosity density, $\rho_{\rm UV}$, at $z>10$.} \label{tab:uvlf}  
\begin{center}    % used for centering table% title of Table    % is used to refer this table in the text
\setlength\tabcolsep{0.5cm}
\begin{tabular}{c c c}
\hline\vspace{0.1cm}%\hline                        % inserts double horizontal lines
$\langle z \rangle $ & $N^{\dagger}$ & $\rho_{\rm UV}$ \\ % table heading
\vspace{0.1cm}
 & (\# of galaxies) & (erg\,s$^{-1}$\,Hz$^{-1}$\,Mpc$^{-3}$) \\
\hline   % inserts single 
\vspace{0.1cm}
$10.375$ & 12 (24) & $(1.2_{-0.3}^{+0.5})\times 10^{25}$  \\
\vspace{0.1cm}
$11.125$ & 5 (12) & $(4.9_{-2.1}^{+3.3})\times 10^{24}$  \\
\vspace{0.1cm}
$12.25$ & 1 (5) & $(0.9_{-0.7}^{+1.0})\times 10^{24}$  \\
\vspace{0.1cm}
$13.75$ & 3 (6) & $(2.5_{-1.4}^{+2.4})\times 10^{24}$  \\
\hline          %inserts single line
\end{tabular} 
\end{center}  
\textbf{Notes.} The number of spectroscopic confirmed sources indicated here is first for those brightward of $-19.75$ mag and then at all luminosities.  47 spectroscopic redshift determinations are available in total.  Compiled redshift measurements are primarily from \citet{RobertsBorsani25} but also include spectroscopic redshift determinations from the JADES DR4 release \cite{CurtisLake25} (goods-s-mediumjwst\_177075: $z=11.118$), from \citet{Donnan26} ($z=13.53$), from \citet{Rodighiero2026} ($z=11.45$), from \citet{Weibel25} (52799: $z=10.32$ and 72355: $z=10.42$), from \citet{Egami2025jwst} (8060\_203922: $z=10.50$), and from \citet{Carnall2024EXCELS} ($z=10.12$).  For this indicative estimate of the luminosity density at $z>10$, only galaxies brighter than $-19.75$ mag are utilized.  The sources to this limiting luminosity (integrated down to $M_{\rm UV} = -17$\,mag) is assumed to provide an indicative estimate of the true luminosity density at $z>10$.  The reported uncertainties represent the Poisson errors based on the galaxy number count in each bin. 
See also \citet{Bouwens2026} for a parallel presentation of this result.%{\bf References.} ??
\end{table}
%%%%%%%%%%%%%%

\begin{figure}[!t]
    \centering
    \includegraphics[width=9.2cm]{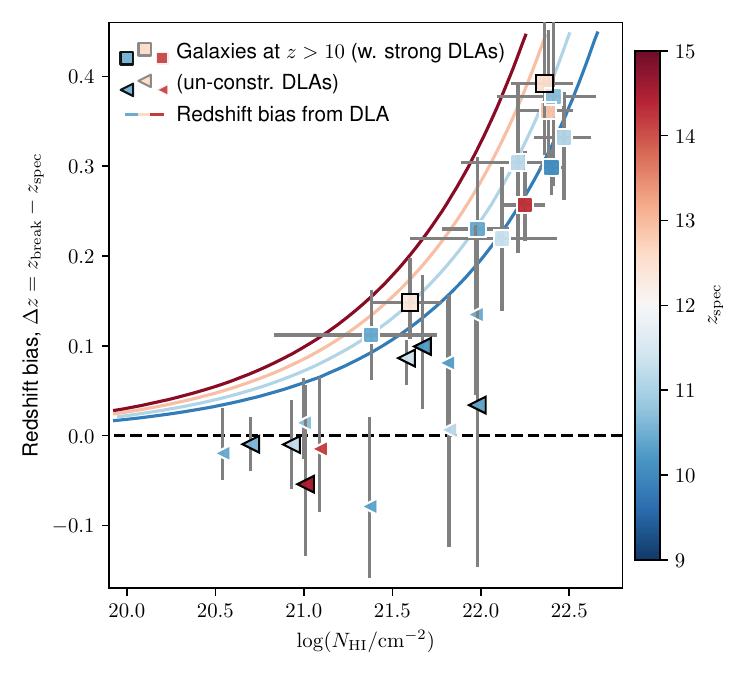}
    \caption{Redshift bias, $\Delta z = z_{\rm break}-z_{\rm line}$, as a function of \hi\ gas column density. The symbol notation follows Fig.~\ref{fig:ksplot}, but here color-coded as a function of spectroscopic redshift, $z_{\rm spec}$. The colored tracks show the evolution of the bias with redshift, resulting in a linearly increasing bias with redshift due to an increasing absorption trough length, measured in terms of observed wavelength, at fixed $N_{\rm HI}$. The dashed line shows where the \lya\ break and spectroscopic line redshifts are equal. }
    \label{fig:zbiasnhi}
\end{figure}

We caution that the redshift bias is further expected to grow linearly with redshift at a fixed \hi\ column density, due to the size of the absorption trough, and hence the offset of the break from the systemic \lya\ redshift, increasing in the observed frame\footnote{This holds for the absolute bias $\Delta z$, whereas the relative redshift bias $\Delta z / (1+z)$ depends only on \hi\ column density and not on redshift itself.}. In Figure~\ref{fig:zbiasnhi}, we show this quantitatively by comparing the redshift bias $z_{\rm break}$ to the derived $N_{\rm HI}$ as a function of redshift. For visual aid we overplot the simulated tracks of $z_{\rm break}$ as a function of $N_{\rm HI}$ from $z=9-15$. It is clear that galaxies with strong DLAs generally follow the expected tracks, observed to increase up to $\Delta z_{\rm break} = 0.4$ for instance in JADES-GS-z12-0. We note, however, that it may be possible in some cases to constrain the \hi\ column density and galaxy redshift directly from the \lya\ wing in spectra with strong DLAs and high S/N based on the shape of the wing itself \citep[e.g.,][]{Witstok25_O3,Pollock26}.
These results emphasize the need for emission-line measurements for robust redshift inferences at $z>10$. Additionally, it highlights the need for photometric or \lya\ break redshift inferences to take into account both the increased prominence and prevalence of DLAs in galaxies at higher redshifts and the redshift-dependent effect on the bias itself.  

\begin{figure*}[!t]
    \centering
    \includegraphics[width=16cm]{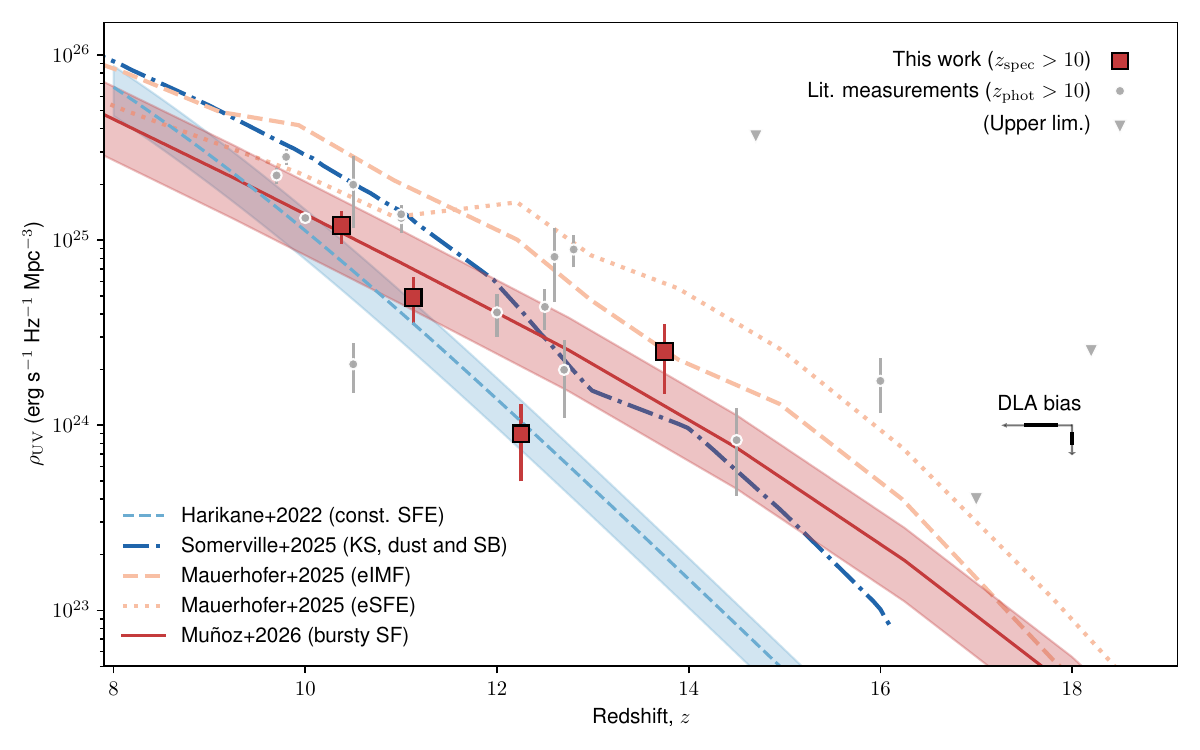}
    \caption{Co-moving cosmic UV luminosity density, $\rho_{\rm UV}$, as a function of redshift for $z>10$. The spectroscopic points derived in this work are shown by the red squares. Select literature measurements derived from \jwst\ photometry are shown for comparison in grey (see Main Text). The mean effect of the redshift-bias on both $\rho_{\rm UV}$ and $z$ due to strong DLAs in the galaxy spectra are indicated by the grey and black arrows, showing the $2\sigma$ and $1\sigma$ distribution interval, respectively. Predictions for the evolution of $\rho_{\rm UV}(z)$ are shown for pre-\jwst\ measurements \citep[blue dashed,][assuming a constant star-formation efficiency]{Harikane22} and for three new models invoking a density-modulated starburst \citep[blue dot-dashed,][]{Somerville25}, an evolving IMF (orange dashed) or SFE \citep[orange dotted,][]{Mauerhofer25}, or including highly stochastic star formation \citep[red solid,][]{Munoz26}.}
    \label{fig:uvlf}
\end{figure*}

The exact shape and normalization of the fundamental UV luminosity function (UVLF) at $z>10$ has been the subject of heavy and ongoing debate since the launch of the \jwst\ \citep[e.g.,][]{Castellano22, Naidu22, Atek23a, Bouwens23, Donnan23, Finkelstein23, Harikane23, Adams24, McLeod24, Robertson24, Whitler25,Weibel25,PerezGonzalez25,Castellano25a}. While there is some discrepancy on the overall functional form, most studies have found converging evidence for a higher co-moving UV luminosity density, $\rho_{\rm UV}$, at $z>10$ than extrapolations from previous, lower-redshift results would infer.

%\begin{figure*}
%\floatbox[{\capbeside\thisfloatsetup{capbesideposition={right,bottom},capbesidewidth=6cm}}]{figure}[\FBwidth]
%{ \caption{Co-moving cosmic UV luminosity density, $\rho_{\rm UV}$, as a function of redshift for $z>10$. The spectroscopic points derived in this work are shown by the red squares. Select literature measurements derived from \jwst\ photometry are shown for comparison in grey (see Main Text). The mean effect of the redshift-bias on both $\rho_{\rm UV}$ and $z$ due to strong DLAs in the galaxy spectra are indicated by the grey and black arrows, showing the $2\sigma$ and $1\sigma$ distribution interval, respectively. Predictions for the evolution of $\rho_{\rm UV}(z)$ are shown for pre-\jwst\ measurements \citep[blue dashed,][assuming a constant star-formation efficiency]{Harikane22} and for three new models invoking a density-modulated starburst \citep[blue dot-dashed,][]{Somerville25}, an evolving IMF (orange dashed) or SFE \citep[orange dotted,][]{Mauerhofer25}, or including highly stochastic star formation \citep[red solid,][]{Munoz26}.}
%\label{fig:uvlf}}
%{\includegraphics[width=11.5cm]{Figs/UVLFz10.pdf}}
%\end{figure*}

To gauge the typical impact of the photometric redshift bias on the $UV$ LF, we made use of the current set of spectroscopic redshifts available above $z>10$.  Most of these are included in the compilation of \citet{RobertsBorsani25} but we also include other more recent spectroscopic redshift determinations, including those from the JADES DR4 release \cite{CurtisLake25} (goods-s-mediumjwst\_177075: $z=11.118$), from \citet{Donnan26} ($z=13.53$), from \citet{Rodighiero2026} ($z=11.45$), from \citet{Weibel25} (52799: $z=10.32$ and 72355: $z=10.42$), from \citet{Egami2025jwst} (8060\_203922: $z=10.50$), and from \citet{Carnall2024EXCELS} ($z=10.12$). 
We then take the luminosities and redshifts of spectroscopically-confirmed galaxies as representative 
of the luminosity density at $z>10$.  Given the high visibility of spectroscopic redshift confirmations at very high redshifts, i.e., $z>11$, vis-a-vis confirmations at lower redshifts, it is reasonable to suppose that $z>11$ are more well sampled from spectroscopy than sources at lower redshifts.  Given the substantial telescope time required to confirm galaxies at lower luminosities, we only consider spectroscopic confirmations brightward of $-19.75$ mag and treat those as representative in deriving an indicative evolution of the $UV$ luminosity density.  These are not sampled by requiring a similar strict S/N threshold in the rest-frame UV as our work, which was essential to robustly model the \lya\ absorption wing.

It has been conventional to consider determinations of the $UV$ luminosity density integrated down to $-17$ mag, which is 2.75 mag fainter.  Assuming that the faint-end slope $\alpha$ has an approximate value of $\sim-2.2$ at $z\geq 10$ as implied by the results of \citet{Bouwens2022_hff,Leung2023,Chemerynska2025,Whitler2025} and there is minimal evolution in the faint-end slope $\alpha$ results from $z\sim14$ to $z\sim10$ as suggested by the \citet{Chemerynska2025} results, this results in a $\sim$6$\times$ larger luminosity density at $z>10$ than to $-19.75$ mag.  If we normalize the results to match the median $UV$ luminosity densities derived by \citet{Donnan24,Weibel25,Willott24,Whitler25} at $z\sim10$, this translates into requiring an assumed 50\% completeness for bright spectroscopic probes of $z>10$ galaxies over 660 arcmin$^2$, which is the approximate area of the PRIMER+CEERS and two GOODS fields, which has been where NIRSpec follow-up efforts have primarily focused.  If the actual spectroscopic completeness is lower or higher than this fiducial value, it would translate in systematically higher or lower luminosity densities, respectively, but the purpose of this exercise is to provide some indicative values and are similar in spirit to what \citet{Harikane23,Harikane25} and \citet{Bouwens23} have done on the basis of prior spectroscopic redshift and photometric compilations, respectively.  In deriving luminosity densities for the $z>10$ universe, we consider four bins in redshift: 10.00-10.75, 10.75-11.50, 11.50 - 13.00, and 13.00-14.50. 

The results are summarized in Table~\ref{tab:uvlf} and shown in Fig.~\ref{fig:uvlf}.  These results are also presented in a review article \citep{Bouwens2026}, while the present manuscript provides the full technical development.  For reference, we show one pre-\jwst\ prediction for the redshift evolution of the UVLF, extrapolated to $z>10$ and assuming a constant star-formation efficiency \citep{Harikane22}. Like the majority of past \jwst\ measurements, we find a modest excess in $\rho_{\rm UV}$ at $z>10$ compared to this model, particularly at $z\sim13.75$. More updated frameworks, including the density-modulated star formation efficiency (DMSFE) model presented in \citet{Somerville25},  the more stochastic star formation included in \citet{Munoz26}, and the {\tt SPHINX} high-resolution radiation-hydrodynamics simulation included into the semi-analytical model {\tt DELPHI} \citep{Mauerhofer25}, are instead able to more accurately reproduce the cosmic-average observations. We caution that this is only the case for the specific set of models in \citet{Somerville25}, where for instance the cloud-fd models systematically overestimate our results by a large margin. 

For comparison, we also compile several photometric estimates of $\rho_{\rm UV}$ at $z>10$ from the literature \citep{Bouwens23,Donnan23,Harikane23,Chemerynska24,McLeod24,Whitler25,Weibel25}. Overall, these results are found to agree well with the spectroscopic results and the updated models presented here. We note though some excess values have been photometrically reported, in particular at $z\geq 16$. We caution, however, that no galaxies have been spectroscopically confirmed at this epoch yet. We derive the impact on the photometrically-derived $\rho_{\rm UV}$ by updating the $M_{\rm UV}$'s with the average $\Delta z_{\rm phot}$ bias. While this has some measurable effect on the redshift, as shown by the arrows in Figure~\ref{fig:uvlf}, the change in $M_{\rm UV}$ is only a few percent. We thus surmise that strong DLAs observed in a large fraction of high-redshift galaxies only introduce a marginal redshift bias on the photometrically derived  $\rho_{\rm UV}$, insufficient to explain their apparent excess UV-brightness.   
% Include also plot with actual UV LF measurements? (phi(MUV)) vs MUV? 

\section{Summary \& Outlook}
\label{sec:conc}

In this work, we presented a detailed characterization of a representative sample of 25 galaxies brighter than $M_{\rm UV} = -18$\,mag, that have all been spectroscopically confirmed at $z>10$ with \jwst. These galaxies represent the brightest sources detected to date within the first 500\,Myr of cosmic time. The compiled galaxies are required to have sufficient S/N at rest-frame UV wavelengths to properly constrain the \lya\ damping wings \citep[see also the companion paper by][]{Pollock26}. The main focus of this work was to understand the impact of strong DLA absorption features, which are observed to be prevalent and prominent at these redshifts, on the redshift estimates and the apparent overabundance, UV-brightness, and prominent nebular emission lines of galaxies at this epoch.  

We modeled the \lya\ damping wings of the sample galaxies using three different iterations; one providing the best-fit local \hi\ gas column density, $N_{\rm HI}$, and neutral hydrogen fraction in the IGM, $x_{\rm HI}$, at the redshift of the galaxies. The other were fixed to model the IGM only, to assess the potential redshift bias a strong DLA could introduce when not properly accounting for this in the redshift-inferences, and the third was set to only include an ISM component to determine the maximum upper bound on $N_{\rm HI}$ in the line of sight. We found that including an IGM-only model to derive the redshift from the \lya\ break could overestimate the actual spectroscopic emission-line measured redshift up to $\Delta z = 0.4$ ($\Delta z = 0.14$ on average). This redshift bias was even more severe for photometric inferences that could overestimate the redshift by $\Delta z > 1$ in case of very prominent DLAs \citep[see also][]{Witstok25_O3}, and on average by $\Delta z = 0.39$. We further modeled the \lya\ optical depth as a function of redshift and $N_{\rm HI}$, and demonstrated that the \lya\ break redshift bias clearly increases with both parameters. However, we found that the DLA redshift bias had a marginal effect on the excess UV luminosity function at $z>10$. 

To understand which physical conditions might instead drive this excess UV-bright population of galaxies, we first investigated the connection between the dense local \hi\ gas and the absolute UV brightness, representing also the dark matter halo mass of each source. The former serves as the fundamental fuel of gas available for star formation before it turns molecular. We found no apparent correlation between these quantities, noting that low column-density absorbers are found among the brightest and faintest of the sources considered here. The same was evident for that connection between the \hi\ gas content and the `burstiness' of the galaxies, here defined as ${\rm SFR_{10\,Myr}}/{\rm SFR_{100\,Myr}}$. This quantity was derived from the star-formation histories of the galaxies inferred via spectro-photometric fitting using {\tt Bagpipes}. However, we recovered a significant relation between the galaxy burstiness and the presence of prominent rest-frame UV nebular emission lines in the spectra \citep[see also][]{RobertsBorsani25}. This supports the hypothesis that these high-ionization lines mainly originate from highly star-forming galaxies caught during a recent, short-lived starburst episode. We also established a new, empirical calibration relating the short-lived ${\rm SFR_{10\,Myr}}$ to the \ciii]\,$\lambda 1909$ line luminosity for galaxies at $z>10$, which provides a more direct measure of the burstiness when compared to the UV continuum luminosity. This calibration is further applicable to the planning of future JWST observations targeting galaxies at this epoch, to estimate the required sensitivity to detect at least one of the prominent rest-frame UV nebular lines for accurate redshift determinations. 

We then investigated the location of the sample galaxies on an equivalent `Kennicutt-Schmidt' plane, comparing their local \hi\ gas surface densities to their UV-derived SFR surface densities \citep[first introduced in][]{Pollock26}. We observed a continuous distribution in the location of galaxies with the most prominent recent starbursts, with ${\rm SFR_{10\,Myr}}/{\rm SFR_{100\,Myr}}>3$ across a range of offsets from the local Kennicutt-Schmidt plane indicating a large variation in the efficiency of star formation or gas depletion timescales. Comparing the observations with predictions from hydrodynamical simulations \citep{Lagos25,Shen26} showed that the theoretical models were generally able to reproduce the range in local \hi\ gas surface densities, though severely underestimating the SFR surface densities. From the observations we surmised that these particular galaxies at $z>10$ likely cycle through a phase of rapid gas consumption to star formation, requiring rapid replenishment of continuously infalling IGM gas, such that this external process is the main driver of the starburst episodes in the first galaxies. Finally, we emphasized the observation that the sample galaxies showed a continuous distribution in the UV-based SFR surface densities as well, alleviating the proposed dichotomy in galaxy sizes as previously reported for these redshifts \citep[e.g.,][]{Harikane25,Naidu25}. 

In conclusion, our results suggested that enhanced star formation efficiencies, short gas depletion timescales, and highly stochastic starbursts are in combination the most likely physical origin of the larger population of more UV-bright galaxies at $z>10$. We note, however, that a subset of galaxies may have artificially boosted their apparent stellar UV luminosity due to dominant nebular free-bound or two-photon continuum emission \citep[see also][]{Cameron24,Katz25}. This effect has been found to prevail in $\sim 5-30\%$ of the total galaxy population at $z=5-10$ \citep{Trussler26}, and can potentially enhance the UV magnitude by 2-3\,mag \citep{Reumert26}. The present observations paint a picture where the majority of these luminous galaxies are seen during a short-lived, ultra-bright burst with prominent rest-frame UV emission lines driven by rapid gas depletion and accretion on short $\sim 20$\,Myr timescales and embedded in nascent, dense \hi~gas reservoirs.

\begin{acknowledgements}
% We would like to thank X, Y, and Z. 

KEH acknowledges support from the Independent Research Fund Denmark (DFF) under grant 5251-00009B and co-funding by the European Union (ERC, HEAVYMETAL, 101071865). Views and opinions expressed are, however, those of the authors only and do not necessarily reflect those of the European Union or the European Research Council. Neither the European Union nor the granting authority can be held responsible for them. 
The Cosmic Dawn Center (DAWN) is funded by the Danish National Research Foundation under grant DNRF140. The data products presented herein were retrieved from the Dawn \jwst Archive (DJA). DJA is an initiative of the Cosmic Dawn Center (DAWN), which is funded by the Danish National Research Foundation under grant DNRF140. This work has received funding from the Swiss State Secretariat for Education, Research and Innovation (SERI) under contract number MB22.00072, as well as from the Swiss National Science Foundation (SNSF) through project grants 200020\_207349 and 2000-1-243073.

Software used in developing this work includes: \texttt{matplotlib} \citep{matplotlib}, \texttt{numpy} \citep{numpy}, \texttt{scipy} \citep{scipy}, \texttt{TOPCAT} \citep{topcat}, and \texttt{Astropy} \citep{astropy}.

This work is based on observations made with the NASA/ESA/CSA James Webb Space Telescope. The data were obtained from the Mikulski Archive for Space Telescopes at the Space Telescope Science Institute, which is operated by the Association of Universities for Research in Astronomy, Inc., under NASA contract NAS 5-03127 for \jwst. 
\end{acknowledgements}

\bibliographystyle{aa}
\bibliography{ref.bib}

\clearpage
\newpage

\begin{appendix}

\section{Establishing \ciii$]\lambda 1909$ as a proxy for bursty star formation} \label{sec:appA}

The \ciii]\,$\lambda\lambda 1907,1909$ doublet transition (hence simply \ciii]\,$\lambda 1909$) is typically one of the most prominent emission line features in high-redshift galaxies seen by \jwst\ \citep[e.g.,][]{RobertsBorsani24,Pollock26}. At lower redshifts, $z\sim 2-3$, it has been shown to correlate with the \lya\ equivalent width \citep{Shapley03,Stark14} and is increasingly detected in more metal-poor, highly ionized galaxies \citep{Erb10,Stark15}. This makes \ciii]\,$\lambda 1909$ one of the primary lines to target at high redshifts, both to accurately pin-point the redshift and to understand their physical properties in terms of metallicity, ionization, and short-lived SFR, given also that \lya\ is decreasingly detected at $z\gtrsim 6$ due to increasing neutral hydrogen fraction of the IGM \citep[e.g.,][]{MiraldaEscude98,McQuinn07}. Indeed, in the sample of galaxies at $z>10$ studied here, the \ciii]\,$\lambda 1909$ line is the most commonly detected.

To fully optimize the use of \ciii]\,$\lambda 1909$ at these early cosmic times, we need to understand its origin. Theoretically, the upper state of the \ciii]\,$\lambda 1909$ transition is mainly populated by photo-ionization of massive stars in sub-solar ($Z\lesssim 0.5\,Z_\odot$) systems with young stellar populations. This implies that the total emission should trace the current star formation on short, $\sim 3-10$\,Myr timescales \citep[e.g.,][]{Erb10,Stark14,Stark15}. Establishing such a relation would serve as a much more direct measure of the `bursty' star formation, comparing the \ciii]\,$\lambda 1909$ line luminosity to the UV continuum. This has so far mainly been derived based on the star-formation history from more complex SED fitting \citep[e.g.,][]{Endsley25,Cole25,Kokorev25,CarvajalBohorquez25,Simmonds25}.  

\begin{figure}[!t]
    \centering
    \includegraphics[width=9cm]{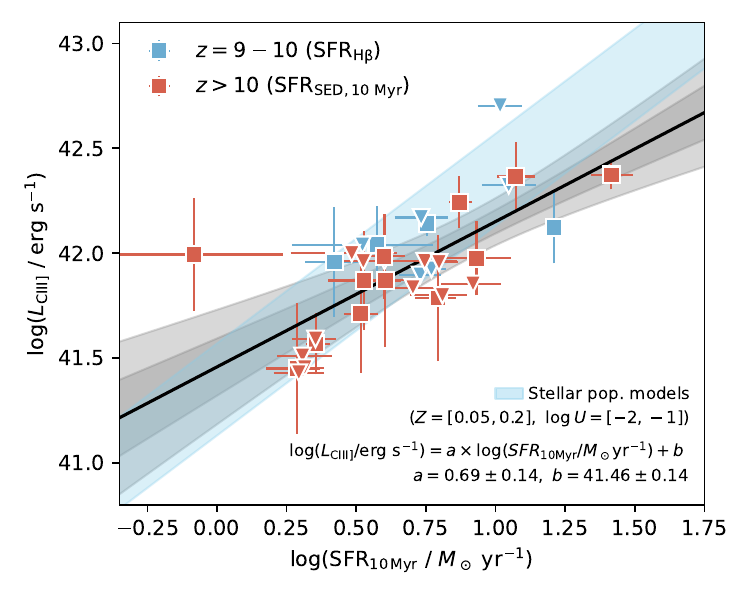}
    \caption{The high-redshift $L_{\rm CIII]}-{\rm SFR_{10~Myr}}$ calibration. The data include measurements of ${\rm SFR_{10~Myr}}$ from H$\beta$ (blue) from the sample of galaxies at $z=9-10$ in \citet{Pollock25} and the compiled sample of galaxies presented here with ${\rm SFR_{10~Myr}}$ from the SED model. The best-fit log-linear relation is shown by the black lines and listed in the bottom right, and the associated $1\sigma$ and $2\sigma$ confidence intervals on the relation are shown by the dark- and light-grey shaded regions, respectively. Predictions for $L_{\rm CIII]}$ are shown for synthetic spectra created with {\tt Bagpipes} for stellar populations models with $Z=[0.05,0.2]$, $\log U = [-2,-1]$, $M_\star = 10^{8}\,M_\odot$, and $A_V = 0.1$}
    \label{fig:ciiisfr}
\end{figure}

In Figure~\ref{fig:ciiisfr}, we compare the SED-derived SFRs over the past 10\,Myr for the sample galaxies studied here to the measured \ciii]\,$\lambda 1909$ line luminosities, $L_{\rm CIII]}$, or upper bounds. We also include the sample from \citet{Pollock25} at $z=9-10$, which has a more direct estimate of SFR$_{\rm 10\,Myr}$ from the H$\beta$ line luminosity at almost comparable redshifts. We find that these two samples are well in agreement, and derive a log-linear relation of $\log(L_{\rm CIII]} / {\rm erg\,s^{-1}}) = a\times \log ({\rm SFR}_{\rm 10\,Myr}/M_\odot\,{\rm yr}^{-1}) + b$, with a best-fit slope and intercept $a=0.69\pm 0.14$ and $b=41.46\pm 0.14$, respectively. 
%\footnote{The authors would like to mention that they asked {\tt ChatGPT} the following prompt: {\tt What is the relation between \ciii]-1909 line luminosity and SFR-10\,Myr for a typical galaxy at $z=10$?} This came up with the mean relation plotted in Fig.~\ref{fig:ciiisfr} without considering the data as input.}. 
For comparison, we also create synthetic galaxy spectra from stellar population models with metallicities in the range $Z=[0.05,0.2]$ and ionization parameters $\log U = [-2,-1]$ using {\tt Bagpipes}, at a fixed stellar mass of $M_\star = 10^{8}\,M_\odot$ and dust attenuation, $A_V = 0.1$\,mag and integrating the SFR over the past 10\,Myr. These are set to represent the average galaxy population at these redshifts. From these mock spectra we measure the range of output \ciii]\,$\lambda 1909$ line luminosities, shown by the blue-shaded region in Fig.~\ref{fig:ciiisfr}. This is in remarkable agreement with the empirically derived relation, though suggesting a slightly steeper slope near unity between $L_{\rm CIII]}$ and ${\rm SFR}_{\rm 10\,Myr}$.

These relations provide useful, empirical starting points for estimating the required sensitivity to detect at least one rest-frame UV nebular emission line at $z>10$ with \jwst/NIRSpec for accurate redshift determinations given an approximate $M_{\rm UV}$ and an estimate of the burstiness of the galaxy based on the SED.   

\end{appendix}

\end{document}